\documentclass{aastex61}

\submitjournal{ApJS}
\shorttitle{Lorentz-Invariance Violation with Crab Pulsar}
\shortauthors{MAGIC Collaboration}
\usepackage{epstopdf,xspace,color}
\usepackage[T1]{fontenc} 
\usepackage{amsmath,amssymb,relsize}
\usepackage{booktabs,verbatim}

\newcommand*\diff{\mathop{}\!\mathrm{d}} 

\newcommand{\avg}[1]{\left< #1 \right>} 
\newcommand{\sigmap}{\sigma_\mathrm{P2}} 
\newcommand{\hsigmap}{\widehat{\sigma}_\mathrm{P2}} 
\newcommand{\phip}{\phi_\mathrm{P2}} 
\newcommand{\hphip}{\widehat{\phi}_\mathrm{P2}} 
\newcommand{\epr}{E^\prime} 
\newcommand{\phipr}{\phi^\prime} 
\newcommand{\bonu}{{\boldsymbol \nu}} 

\newcommand{\ptwo}{P2\xspace}
\newcommand{\pone}{P1\xspace}
\newcommand{\eqg}{E_\mathrm{QG}\xspace}
\newcommand{\eqgn}{E_{\mathrm{QG}_n}\xspace}
\newcommand{\eqgone}{E_{\mathrm{QG}_1}\xspace}
\newcommand{\eqgtwo}{E_{\mathrm{QG}_2}\xspace}

\begin{document}
\title{Constraining Lorentz invariance violation using the Crab Pulsar emission observed up to TeV energies by MAGIC}

\correspondingauthor{Markus Gaug}
\email{markus.gaug@uab.cat}

\AuthorCallLimit=200
\collaboration{MAGIC Collaboration:}

\author{M.~L.~Ahnen}
\affiliation{ETH Zurich, CH-8093 Zurich, Switzerland}
\author{S.~Ansoldi}
\affiliation{Universit{\`a} di Udine, and INFN Trieste, I-33100 Udine, Italy}
\affiliation{Japanese MAGIC Consortium: ICRR, The University of Tokyo, 277-8582 Chiba, Department of Physics, Kyoto University, 606-8502 Kyoto, Tokai University, 259-1292 Kanagawa, The University of Tokushima, 770-8502 Tokushima, Japan}
\author{L.~A.~Antonelli}
\affiliation{INAF - National Institute for Astrophysics, I-00136 Rome, Italy}
\author{C.~Arcaro}
\affiliation{Universit{\`a} di Padova and INFN, I-35131 Padova, Italy}
\author{A.~Babi\'c}
\affiliation{Croatian MAGIC Consortium: University of Rijeka, 51000 Rijeka, University of Split - FESB, 21000 Split, University of Zagreb - FER, 10000 Zagreb, University of Osijek, 31000 Osijek and Rudjer Boskovic Institute, 10000 Zagreb, Croatia}
\author{B.~Banerjee}
\affiliation{Saha Institute of Nuclear Physics, HBNI, 1/AF Bidhannagar, Salt Lake, Sector-1, Kolkata 700064, India}
\author{P.~Bangale}
\affiliation{Max-Planck-Institut f{\"u}r Physik, D-80805 M{\"u}nchen, Germany}
\author{U.~Barres de Almeida}
\affiliation{Max-Planck-Institut f{\"u}r Physik, D-80805 M{\"u}nchen, Germany}
\author{J.~A.~Barrio}
\affiliation{Universidad Complutense, E-28040 Madrid, Spain}
\author{J.~Becerra Gonz\'alez}
\affiliation{Inst. de Astrof\'isica de Canarias, E-38200 La Laguna and Universidad de La Laguna, Dpto. Astrof\'isica, E-38206 La Laguna, Tenerife, Spain}
% ...
\begin{abstract}
 Spontaneous breaking of Lorentz symmetry at energies on the order of the Planck energy or lower
is predicted by many quantum gravity theories, implying non-trivial dispersion relations for the photon in vacuum.
Consequently, gamma-rays of different energies, emitted simultaneously from astrophysical
sources, could accumulate measurable differences in their time of flight until 
they reach the Earth. Such tests have been carried out in the past using fast variations of gamma-ray flux from pulsars, and more recently from active galactic nuclei and 
gamma-ray bursts. 
We present new constraints studying the gamma-ray emission of the galactic Crab Pulsar,
recently observed up to TeV energies by the MAGIC collaboration.
A profile likelihood analysis of pulsar events reconstructed for energies above 400~GeV 
finds no significant variation in arrival time as their energy increases. 
Ninety-five percent~CL limits are obtained on the effective Lorentz invariance violating energy scale at the level of
$\eqgone > 5.5\cdot 10^{17}$~GeV ($4.5\cdot 10^{17}$~GeV) for a linear, and $\eqgtwo > 5.9\cdot 10^{10}$~GeV ($5.3\cdot 10^{10}$~GeV) for a quadratic scenario,
for the subluminal and the superluminal cases, respectively. 
A substantial part of this study is dedicated to calibration of the test statistic, with respect to bias and coverage properties.
Moreover, the limits take into account systematic uncertainties, found to worsen the statistical limits by about 36--42\%. 
Our constraints would have resulted much more competitive if the intrinsic pulse shape of the pulsar between 200~GeV and 400~GeV 
was understood in sufficient detail and allowed inclusion of events well below 400~GeV.
\end{abstract}




\section{Introduction}\label{sec.intro}

Common  models of quantum gravity (QG) \citep{rovelli04} try to 
combine Einstein's framework of gravitation with modern quantum field theory, introducing microscopic granular structure and probabilistic dynamics of space-time. 
Although none of these scenarios is currently universally accepted, most of them~\citep{Kostelecky1989,lqgoptics,Douglas2001,Burgess2002,Magueijo:2002,HamedArkani2004,Horava:2009} 
predict spontaneous violation of the Lorentz invariance (LIV). 
This can lead to a non-trivial, i.e. energy-dependent dispersion relation of the photon in vacuum
and birefringence, as well as an anisotropy of the vacuum. 
At lower energies, the modified dispersion relation can be parameterized by an effective QG energy scale ($\eqg$), which can be on the order of the Planck scale
($E_{\mathrm{Pl}} = \sqrt{\hbar	c^5 / G} \approx 1.22 \cdot 10^{19}$~GeV)
or lower. QG effects are then largely suppressed, but can manifest themselves if photons of different energy travel very large distances and
hence accumulate tiny delays that yield potentially measurable effects~\citep{amelino2009}.



 

The group velocity of photons of energy $E \ll \eqg$ can then be parameterized as (see, e.g.,~\citet{amelino2009}, Eq.~3): %
\begin{equation}\label{eq.livphotonspeed}
	u_{\gamma}(E) = \frac{\partial E}{\partial p} \approx c \cdot \left[ 1 - \sum_n \xi_n \frac{n+1}{2} \left( \frac{E}{\eqgn}\right)^n \right]  \quad,
\end{equation}
where $c$ is the (Lorentz-invariant) speed of light and $\xi_n$ is the sign of the change: $\xi_n = +1$ for a ``subluminal'' scenario (decreasing photon speed with increasing energy), 
$\xi_n = -1$ for the ``superluminal'' case (increasing photon speed with increasing energy), and $\xi_n =0$ for the case that the $n$th order is forbidden. 
The modified dispersion relation can also be written in terms of coupling constants $f_\gamma^{(n)}$ of the minimal standard model extensions (SME)~\citep{colladay1998}, 
in which case the substitutions: $n \rightarrow n + 2$ and $-f_\gamma^{(n)}/E_\mathrm{Pl}^{(n-2)} \rightarrow \xi_n/\eqgn$ lead to the form chosen in Eq.~\ref{eq.livphotonspeed} 
(see also Eqs.~15 and 74 of \citet{livtests2005}).
Eq.~\ref{eq.livphotonspeed} neglects terms breaking rotation invariance,
which  would, however, imply some breaking of boost invariance  if they were present (see again~\citet{livtests2005}, chapter 3.1). 
Terms with $n > 0$ produce energy-dependent velocities and are typically considered in time of flight experiments.\footnote{%
  Note that  terms of $n = -1$ and $n = 0$ are also allowed by SME~\protect\citep{colladay1998}, but strongly constrained by Earth-based experiments.
} 
Because odd terms of $n$ violate CPT~\citep{colladay1998}, 
the $n=2$ term may dominate if CPT is conserved. 
From a theoretical point of view, subluminal propagation is equally plausible as superluminal~\citep{amelino2009}; birefringence effects are also 
possible, in which photons show subluminal and superluminal propagation, depending on their circular polarization state~\citep{Kostelecky:2004,Covino:2016}. Nevertheless, 
birefringence has been strongly bound by other means~\citep{gubitosi2009,Goetz:2014,Kislat:2017} and will not be considered in the following. 
Actually, in the framework of an SME approach,
current limits from astrophysical polarization 
measurements constrain variations of the speed of light that are linear with photon energy to several orders of magnitude beyond the Planck scale~\citep{Goetz:2014,Kislat:2017}.\footnote{However, very recently spectral lags have been claimed in both the neutrino and the photon sector~\protect\citep{amelino2017}, which depend linearly  on the photon energy 
and are incompatible with the above constraints.}
However, not all quadratic terms, some of which may be realized without vacuum birefringence~\citep{Kostelecky:2008}, are constrained. 
Therefore, and behind this background, constraining the quadratic term $\eqgtwo$ is now of particular interest.





Exploiting the fast variations of gamma-ray signals from astrophysical sources at cosmological distances to limit LIV 
was first suggested in~\citet{amelino98}. 
So far, flares from active galactic nuclei (AGNs), exploited first by the Major Atmospheric Gamma-ray Imaging Cherenkov system \citep[MAGIC,][]{liv-mkn501} and later by H.E.S.S.~\citep{hessliv2011},  
and the very fast flux variations of gamma-ray bursts (GRBs), observed by FERMI~\citep{fermiliv13}, 
have boosted sensitivities to such energy-dependent delays and achieved astonishingly strict limits on $\eqgone$, of well beyond the
order of the Planck scale~\citep{fermiliv13,Goetz:2014}. 
Both types of sources have been detected at cosmological distances, but their maximum observable energy is increasingly limited due to 
extinction of photons by the extra-galactic background light (EBL)~\citep{dominguez2015}. 
AGN flares, on the other hand, have been observed until energies of several TeV~\citep{liv-mkn501,hessliv2011}, but are closer in distance and show slower rise and fall times than GRBs. 
Obtained limits on LIV are nevertheless competitive, due to the higher energies achieved, particularly for the quadratic term. 
Both types of sources require
a solid emission model 
in order to discard any intrinsic, insufficiently understood, energy-dependent effects on the time of emission, 
which is not yet the case~\citep{bednarekwagner2008}. The effect of this can be mitigated however, through
the observation of sources at different red-shifts. 


Gamma-ray pulsars, albeit being observed many orders of magnitude closer~\citep{2ndfermi} than AGNs or GRBs,  
have the advantage of  precisely timed regular flux oscillations, with periods down
to the order of milliseconds, as well as the fact that they are the only stable (in the sense of periodically emitting) candidate sources 
for astrophysical time of flight tests, prescinding from the need
of target of opportunity alerts. 
Sensitivity to 
LIV can hence be systematically planned and improved using longer observation times. 
An LIV-induced variation of the speed of light would produce a shift in the position of the pulsar peak
in its phaseogram, i.e. the emission as a function of the pulsar rotational phase.  
Moreover, possible intrinsic energy-dependent time delays from the pulsar itself would be observed proportional to its rotational period, while 
LIV induced effects are not, allowing to disentangle between both, once measurements have been carried out over several years~\citep{nepomuk}.

Actually, the first-ever astrophysical limit on LIV 
was obtained from the Crab Pulsar using optical and radio data~\citep{craborig1969}. First limits on LIV using gamma-ray emission from 
the Crab Pulsar were computed from EGRET data up to 2~GeV~\citep{kaaretcrab}, 
and improved by VERITAS using very high energies (VHE) gamma-rays reaching up to 120~GeV~\citep{nepomuk}. 
Recently, the MAGIC collaboration has published the detection of pulsed emission from the Crab Pulsar up to TeV energies~\citep{crabtev}.
We exploit this unique set of data to derive improved limits on the effective QG scale using a profile likelihood approach, calibrated both in terms of bias and coverage, 
and include systematic uncertainties.




This paper is structured as follows: first, 
we introduce the data set taken on the Crab Pulsar, emphasizing information relevant for LIV searches (Section~\ref{sec.pulsar}). 
Second, we perform a basic peak comparison search for signatures of 
LIV in Section~\ref{sec.method1} and subsequently construct the full likelihood in Section~\ref{sec.method2}.   
Several results from applying these methods to data are presented in Section~\ref{sec.results} and limits to LIV are derived.
A thorough calibration of the likelihood using toy Monte Carlo (MC) simulations is performed in Section~\ref{sec.checks} 
and systematic uncertainties discussed in Section~\ref{sec.uncertainties}. The obtained new limits and their implications will be discussed at the end, 
in Sections~\ref{sec.discussion} and~\ref{sec.conclusions}.

\section{The data set}
\label{sec.pulsar}


The Crab Pulsar PSR~J0534+2200, located at the center of the Crab Nebula in the Taurus constellation, is one of the best-studied
pulsars due to its youth, proximity, brightness, and wide spectral coverage~\citep{crabreview}. 
It shows a rotation period of $T \approx 33.7$~ms,  slowly increasing by $\dot{T} = 4.2\times 10^{-13}$.
Its distance is still rather poorly determined~\citep{Trimble1973} and generally stated as $2.0 \pm 0.5$~kpc~\citep{Kaplan2008}.
The Crab Pulsar phaseogram (defined as flux -- or simply count rate -- as a function of the pulsar phase $\phi$) 
shows increased emission in two phase ranges: the main pulse \pone,
which has been used to define the zero phase value, and the inter-pulse \ptwo at $\phi \approx 0.4$. 
The bridge region between \pone and \ptwo also exhibits emission
in optical, X-rays and, as was discovered by MAGIC~\citep{bridge14}, in VHE gamma-rays between 50 and 150~GeV.
The inter-pulse becomes dominant only at the high end of the spectrum~\citep{crabfermi10,veritascrab11,magicpulsar12}.
The energy spectrum of both pulses can be described by simple power-laws from 10~GeV on, and extends to at least 
0.5~TeV for $\pone$ and 1.5~TeV for $\ptwo$, as recently measured by MAGIC~\citep{crabtev}.

Contrary to AGNs and GRBs, the gamma-ray signal from the Crab Pulsar at VHE 
is very background-dominated: generated during a supernova explosion in 1054~AD, this young pulsar lies
at the center of a strong gamma-ray emitter, the expanding Crab Nebula. 
The VHE emission of both cannot be spatially resolved so far.







The Major Atmospheric Gamma-ray Imaging Cherenkov system (MAGIC) is located at the Roque de los
Muchachos observatory ($28.8^{\circ}$N, $17.8^{\circ}$W, 2200~m a.s.l.), in the Canary Island of La Palma, Spain.
During its first five years of operation, the MAGIC system consisted of a single 17~m dish telescope \citep{magicperform}.
In 2009, a second telescope was added 
with identical structure, but including several major
improvements in its reflective surface, camera, and the electronics used for signal processing \citep{perform12}.

Between 2011 and 2012, a major upgrade of the MAGIC system was performed to install a new camera and trigger
system for MAGIC-I, after which the two telescopes became almost identical in their hardware components~\citep{perform13}.
In stereoscopic observation mode, the system reaches a maximum sensitivity of $\sim 0.6$\% of the Crab Nebula flux,
for energies above $\sim 300$~GeV in 50~hr of observation~\citep{perform16}. 
Nevertheless, sensitivity is slightly worse for this source, limited by the strong gamma-ray background from the nebula.


Because the Crab Nebula is the brightest steady source in the VHE gamma-ray sky, it is considered a calibration source 
for this energy regime and regularly observed for performance tests. 
The MAGIC telescopes have collected more than 1000~hr of total observation time, taken in every possible
hardware configuration during the past 12~years of operation  (a detailed summary of the employed data set can be 
found in Appendix~\ref{app.datasamples}). 
In total, the MAGIC telescopes
collected $3080 \pm 460$ excess events from the 
 \ptwo region, out of which $544 \pm 92$ had reconstructed energies above 400~GeV.
Moreover, MAGIC was able to confirm that there is a significant difference between the steepness
in the spectrum of the Crab main and inter-pulse. 
We use these very same data to perform tests on LIV,
but select 
only 
events close to \ptwo because they will allow us to reach the highest sensitivity, due to the higher reach in energy while keeping the analysis simple enough. 
For more detailed information about the employed data, we refer the reader to the detection paper~\citep{crabtev}.








\section{Peak comparison method}
\label{sec.method1}

We first apply a straightforward method that compares the differences in mean fitted pulse positions at different energies (employed in previous LIV 
searches from Crab Pulsar data \citep[see e.g.][]{nepomuk}), 
and later a more sophisticated likelihood approach. 
Other possible methods, as the so-called PairView or Sharpness-maximization approaches~\citep{fermiliv13}, have not been exploited in this study.

A simple method to search for energy-dependent delays or advancements in pulse arrival time consists of a direct comparison of peak positions of a pulse.



QG effects predict an average phase delay between photons of mean energies $E_\mathrm{l}$ and $E_\mathrm{h}$ of: 
\begin{equation}\label{eq.timedelay}
  \Delta \phi = \frac{d_\mathrm{Crab}}{c\;P_\mathrm{Crab}} \cdot  \xi_n \frac{n+1}{2} \frac{E^n_\mathrm{h}-E^n_\mathrm{l}}{\eqgn^n}  \qquad,
\end{equation}
\noindent
where $d_\mathrm{Crab}$ is the pulsar distance, $c$ the Lorentz-invariant speed of light, $P_\mathrm{Crab}$ the pulsar period, 
and $E_\mathrm{l}$ and $E_\mathrm{h}$ are the mean energies of two separated energy bands, typically chosen to cover the highest part of the observed spectrum and a distinct lower part. 

Limits to $\eqg$ can then be derived from limits on $\Delta \phi$ according to:  
\begin{eqnarray}
\eqgn & \gtrsim & \left(\xi_n\frac{n+1}{2}\;\frac{d_{\mathrm{Crab}}}{c\;P_{\mathrm{Crab}}} \; \frac{E_\mathrm{h}^n-E_\mathrm{l}^n}{\Delta\phi} \right)^{1/n} \quad.
\label{eq.livlimits}
\end{eqnarray}

\begin{figure}[ht!]
\centering
\includegraphics[width=0.99\textwidth]{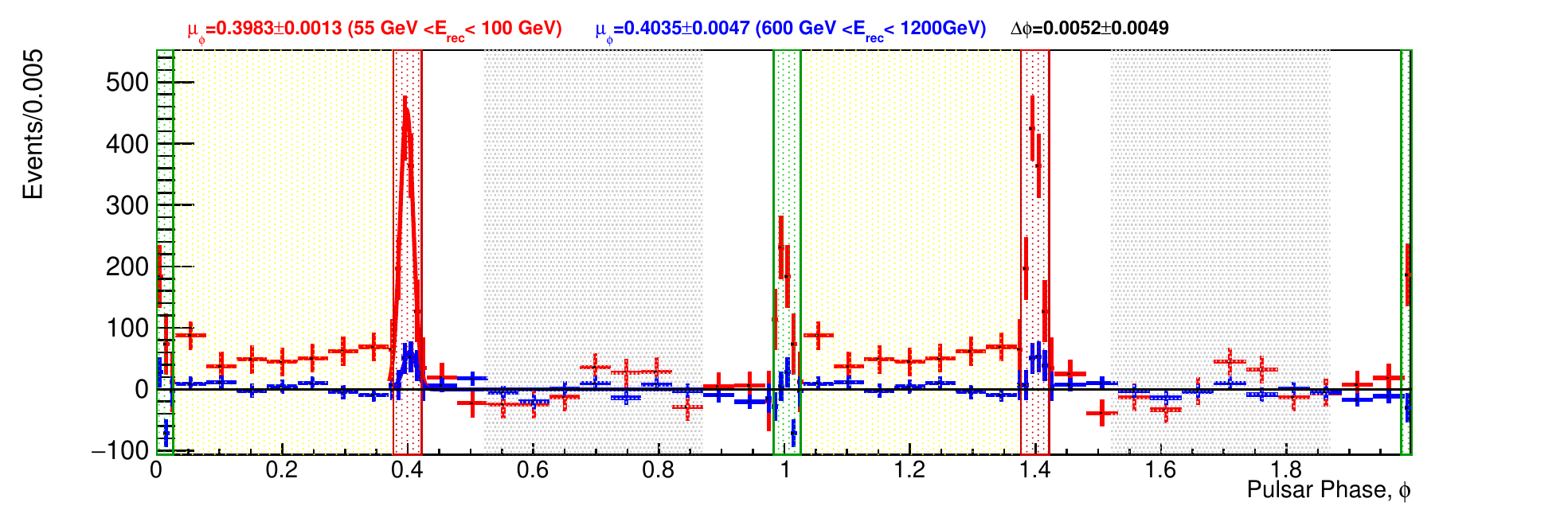} 
\includegraphics[width=0.99\textwidth]{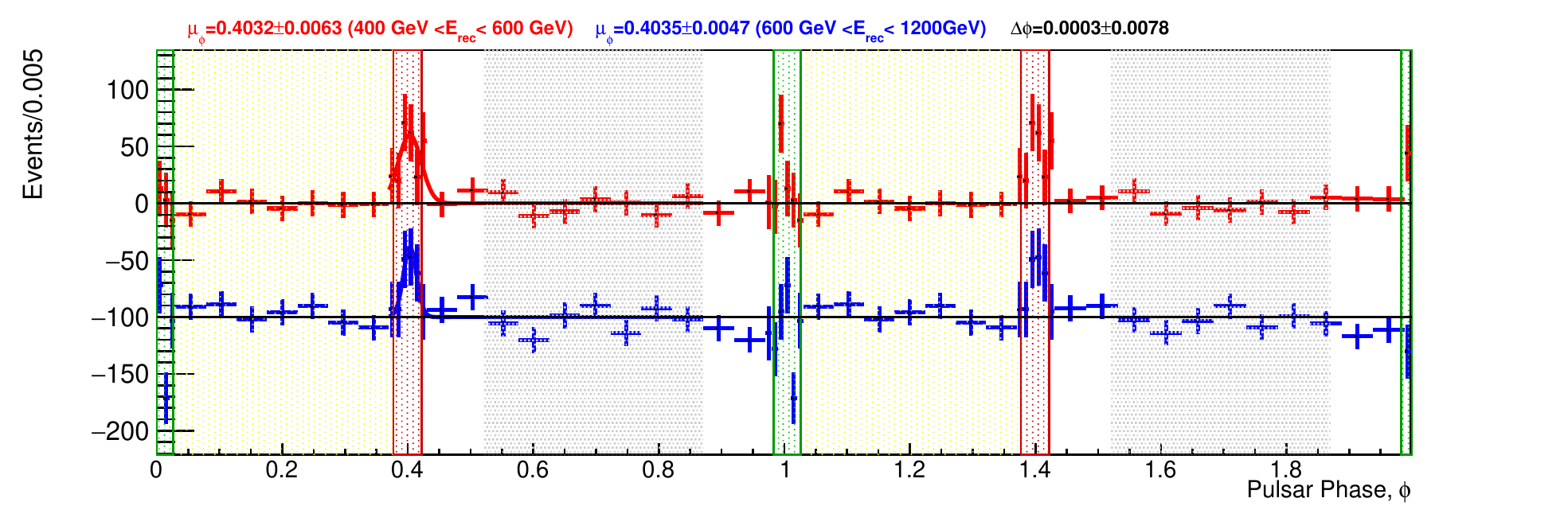} 
\caption{Crab Pulsar folded light curves (two full phases), fitted with a Gaussian pulse shape plus flat background model for the \ptwo inter-pulse. 
The main and inter-pulse regions are binned with a width of 0.005 phases, while the intermediate ranges use coarser bins. 
The bin width of the pulse regions are chosen to fit a reasonable number of bins into the region, 
while the coarse bins are selected to fit an integer number of bins into each region. 
All bin widths have been chosen a priori, independently of the fitting results.  The likelihood fit assumes Poissonian 
fluctuations for the predicted sum of background plus pulse shape from phase 0.37 through 0.87. The background is subtracted here for display only, 
but is included in the fit.
 Two distant energy ranges are shown: between 55 and~100~GeV (top) or 400 and 600~GeV (bottom) in red, and above 600~GeV (both figures) in blue. 
The last one is artificially offset by 100 counts for better visibility.
  The traditional OFF-region is underlaid with a gray area, the \pone region with green, and the \ptwo region with red~\protect\citep{crabtev}. 
The bridge region~\protect\citep{bridge14} between \pone and \ptwo is underlaid with yellow.}
\label{fig.pulsarLCs}
\end{figure}

We use Eq.~\ref{eq.livlimits} to compare the highest possible energy band with sufficient statistics, i.e. from  
 600 to 1200~GeV with two lower bands: 
in one case, they span from the analysis threshold of 55 to 100~GeV, while in the second scenario, 
the lower band limits itself to the data published in~\citet{crabtev} and ranges from 400 to 600~GeV. 
The second choice is motivated by the fact that the detection of pulsar emission at such high energies can  
hardly be reconciled with the traditional interpretation of pulsar emission through the synchro-curvature process at lower energies.  
Such a couple of high-energy bands would hence not be affected by a change of the emission mechanism, if such a change happens below $\sim$400~GeV, 
albeit at the price of a worse limit on LIV (see Figure~\ref{fig.pulsarLCs}).


The mean energies of the selected bands of reconstructed energy, 
$E_\mathrm{l} \sim$~75~GeV and 465~GeV, respectively, for the two low-energy bands, and $E_\mathrm{h} \sim$~770~GeV for the high-energy one,
have been found by MC simulations of the energy spectrum of $\ptwo$~\citep{crabtev}, weighted with the correct exposure of the different samples.

\begin{table}[h!]
\centering
\begin{tabular}{ccccc}
\toprule
Nuisance   &     Result     &      Result    &   Result    &  Result \\
 Parameter & (55--100~GeV)  & (400--600~GeV) &  (600--1200~GeV)  &  (400--1200~GeV) \\
\midrule
\multicolumn{1}{c}{\textit{Gaussian Pulse Shape}} &         &                &    & \\
\midrule
$\hphip$   & {\normalsize $0.398 \pm 0.001  $}  & {\normalsize $0.403 \pm 0.006 $}  & {\normalsize $0.404 \pm 0.005 $} & {\normalsize $0.403 \pm 0.004 $}\\  
$\hsigmap$ & {\normalsize $0.011 \pm 0.001  $}  & {\normalsize $0.018 \pm 0.007  $} & {\normalsize $0.011 \pm 0.005 $} & {\normalsize $0.015 \pm 0.005 $}\\  
$\chi^2/$NDF &  1.15  &                0.75             &  0.90    &   1.06    \\\midrule                                                                   
\multicolumn{1}{c}{\textit{Lorentzian Pulse Shape}}&     &                      &       & \\                                                                        
\midrule                                                                                                                                     
$\hphip$   & {\normalsize $0.399 \pm 0.001  $}  & {\normalsize $0.401 \pm 0.005 $} & {\normalsize $0.403 \pm 0.006 $} & {\normalsize $0.402 \pm 0.004 $} \\  
${\widehat{\gamma}_\mathrm{P2}}$ & {\normalsize $0.010 \pm 0.002  $}  & {\normalsize $0.02  \pm 0.01 $}  & {\normalsize $0.012 \pm 0.008 $} & {\normalsize $0.014 \pm 0.007 $} \\  
$\chi^2/$NDF &  1.14  &                0.73             &  0.89   &  0.97 \\
\bottomrule
\end{tabular}
\caption{Obtained fit values from the Peak Comparison Method. 
The uncertainties are statistical only. The entries labeled $\phi_\mathrm{P2}$ denote the Gaussian or Lorentzian mean, while $\sigma_\mathrm{P2}$ denotes the Gaussian sigma and 
$\gamma_\mathrm{P2}$ the Lorentzian half width at half-maximum. The last column shows results for the full range from 400~GeV--1200~GeV and is not used for the peak comparison, but 
can be compared to the later results in Table~\protect\ref{tab.MLnuisance}.
\label{tab.peaknuisance}}
\end{table}

The pulses are fitted using the method of maximizing a Poissonian likelihood whose mean is parameterized by one Gaussian over a constant background \citep[see][]{magicpulsar12}. 

A simulated signal of two half-Gaussians of different width, joined at the peak, was also tested, but the $\chi^2/\mathrm{NDF}$ of the fits did not improve in any of the tested energy bands.
The fit positions $0.3983 \pm 0.0013_{\text{ stat.}}$ and $0.4032 \pm 0.0063_{\text{ stat.}}$  of the \ptwo peaks in the phaseogram are obtained for the 
two lower-energy bands \citep[compatible with][]{magicpulsar12}, 
and $0.4035 \pm 0.0047_{\text{ stat.}}$ for the high-energy band, compatible with~\citet{crabtev}. 
A Lorentzian pulse shape is also tested, 
yielding marginally better $\chi^2/\mathrm{NDF}$ and compatible results for the pulse positions. See Table~\ref{tab.peaknuisance} for the obtained results.

The delay in the arrival phase between these two energy ranges is then
$\Delta \phi_{\mathrm{P2}} = 0.0052 \pm 0.0049_{\mathrm{stat}} \pm 0.003_{\mathrm{syst}}$ (between 75~GeV and 770~GeV) and 
$\Delta \phi_{\mathrm{P2}} = 0.0003 \pm 0.0078_{\mathrm{stat}} \pm 0.0030_{\mathrm{syst}}$ (between 465~GeV and 770~GeV), both
 compatible with no delay. However, they also show the trend, observed at lower energies, of the mean pulse peak positions slowly shifting toward higher phases, as their energy 
increases \citep[see e.g.][]{crabfermi10,magicpulsar12}. 
The systematic term contains additional uncertainties due to the phase binning and the differences obtained when choosing a Lorentzian light curve model
or asymmetric widths \citep[see also][]{magicpulsar12}. 
The derived 95\% CL limits on an LIV-induced linear and quadratic phase delay are shown in Table~\ref{tab.simplelimits}.



\begin{table}[h!]
\centering
\begin{tabular}{ccc}
\toprule
       &  55--100~GeV & 400--600~GeV  \\
Case  &   versus  &  versus  \\
      & 600--1200~GeV  & 600--1200~GeV \\
\midrule
& \multicolumn{2}{c}{$\eqgone$ (GeV)} \\
\midrule
%
%
%
$\xi_1 = +1$  & $2.5 \times 10^{17}$ &  $1.1 \times 10^{17}$  \\
$\xi_1 = -1$  & $6.7 \times 10^{17}$ & $1.1 \times 10^{17}$ \\
\midrule
& \multicolumn{2}{c}{$\eqgtwo$ (GeV)} \\
\midrule
$\xi_2 = +1$  & $1.8 \times 10^{10}$ & $1.4 \times 10^{10}$ \\
$\xi_2 = -1$  & $2.9 \times 10^{10}$ & $1.5 \times 10^{10}$ \\
\bottomrule
\end{tabular}
\caption{Obtained 95\%CL limits from the Peak Comparison Method: the first two lines apply to the linear case (delay and advancement), while the 
last two lines are valid for the quadratic case of LIV (again delay and advancement). The second column represents the limit on the characteristic 
LIV energy scale, obtained by comparing the two distant energy bins 55--100~GeV and 600--1200~GeV, while the last columns shows the 
limits obtained from the two adjacent high-energy bins 400--600~GeV and 600--1200~GeV. \label{tab.simplelimits}}
\end{table}

\clearpage

\section{Maximum likelihood method}
\label{sec.method2}

More sensitive constraints, which exploit the full information of the MAGIC Crab Pulsar data set, can be obtained with
a maximum likelihood (ML) method, first introduced for this kind of search in~\citet{manelmanel} and further elaborated in~\citet{fermiliv13}. 

We define two new parameters for the linear and quadratic LIV effect intensity, respectively: $\lambda_1 \equiv 10^{19}~\text{GeV}/\eqgone$ and
$\lambda_2 \equiv 10^{12}~\text{GeV}/\eqgtwo$. 
The mean phase delay produced by the LIV effect under test is then 
\begin{eqnarray}
\Delta\phi_n &=& c_n \cdot \bigg( \lambda_n \cdot \left(\frac{E}{\mathrm{GeV}}\right) \bigg)^n~\quad, \label{eq.Deltaphi} \\[0.35cm]
     \mathrm{with:} && \nonumber\\[0.35cm]
c_1 &=& \xi_1 \cdot \frac{d_\mathrm{Crab}}{c\cdot P_\mathrm{Crab}} \cdot 10^{-19}  \quad (\mathrm{GeV}^{-1}) \\[0.3cm]
c_2 &=& \xi_2 \cdot \frac{3}{2}\frac{d_\mathrm{Crab}}{c\cdot P_\mathrm{Crab}} \cdot 10^{-24}  \quad (\mathrm{GeV}^{-2})  \qquad,
\end{eqnarray}
 such that a positive (negative) value of $\xi$ indicates a subluminal (superluminal) scenario, and a zero intensity of $\lambda_{1,2}$ stands for
an infinite LIV energy scale $E_{\mathrm{QG}_{1,2}}$. Note that these definitions differ from those employed by~\citet{hessliv2011} and~\citet{fermiliv13}, particularly for 
$\lambda_2$ which is now directly proportional to $1/\eqgtwo$ instead of $1/E^2_{\mathrm{QG}_2}$. 
 Being closer to the constrained quantity of interest, particularly $\eqgtwo$, our definition will allow us to investigate its statistical properties more accurately (see Section~\ref{sec.checks}). 

Using the profile likelihood ratio method~\citep{murphy2000}, 
we define a test statistic $D_n$ for $\lambda_n$ of our pulsar dataset ${\textit{\textbf X}}=\{\epr_i,\phipr_i,k_i\}$, where $\epr_i$ is the reconstructed energy, $\phipr_i$ is the reconstructed 
phase, $k_i$ is the observation period of event $i$,  and  $\bonu$ is a set of nuisance parameters:
\begin{equation}
D_n(\lambda_n|{\textit{\textbf X}}) = -2 \ln \left(\frac{\;\mathcal{L}(\lambda_n;\widehat{\widehat\bonu}(\lambda_n)|{\textit{\textbf X}})}
                                                  {  \mathcal{L}(\widehat{\lambda}_n; \widehat\bonu    |{\textit{\textbf X}})}\right) \quad. \label{eq.lr1}
\end{equation}

Single-hatted parameters $\{\widehat{\lambda}_n,\widehat\bonu\}$ maximize the likelihood, while 
double-hatted parameters $\widehat{\widehat\bonu}$ are those that maximize $\mathcal{L}$ for a given assumption of $\lambda_n$. 

Some care needs to be taken for the cases where $\widehat{\lambda}_n$ comes to lie in an ``unphysical'' region. We need to define as ``unphysical'' all those 
values that cannot be part of a given theory, due to fundamentally different concepts. In our case, this would mean
\textit{negative values of $\widehat{\lambda}_n$ for subluminal theories, i.e. $\xi_n>0$} and 
\textit{positive values of $\widehat{\lambda}_n$ for superluminal theories, i.e. $\xi_n<0$}. Following the recommendation of~\citet{cowan2013}, we adopt an 
alternative test statistic that avoids the formal use of physical boundaries by construction, namely:

\begin{eqnarray}
\tilde{D}_n(\lambda_n|{\textit{\textbf X}}) = \left\{ 
\begin{array}{ll} 
  -2 \ln \left(\frac{\;\mathcal{L}(\lambda_n;\widehat{\widehat\bonu}(\lambda_n)|{\textit{\textbf X}})}
  {  \mathcal{L}(\widehat{\lambda}_n; \widehat\bonu    |{\textit{\textbf X}})}\right) & \quad\mathrm{if} ~ \mathrm{sgn}(\widehat{\lambda}_n) = \mathrm{sgn}(\xi_n) \\[0.4cm]
 -2 \ln \left(\frac{\;\mathcal{L}(\lambda_n;\widehat{\widehat\bonu}(\lambda_n)|{\textit{\textbf X}})}
  {  \mathcal{L}(0; \widehat{\widehat{\bonu}}(0)|{\textit{\textbf X}})} \right)       & \quad\mathrm{otherwise}\quad.
\label{eq.lr}
\end{array} 
\right.
\end{eqnarray}

Such a test statistics allows us to set limits on $\lambda_n$ at a given confidence level. 
A one-sided 95\% CL limit is then determined by the value of $\lambda_n$ at which $\tilde{D}_n \approx 2.705$~\citep{pdg}.

We compute $\mathcal{L}$ in the form of an \textit{extended likelihood}, i.e. the product of the probability density function (PDF) ($\mathcal{P}$) of each
event in our dataset, considered independent among each other, multiplied by the Poissonian probability to obtain the number of observed events, given the hypothesis $\{\lambda_n;\bonu\}$ 
(see, e.g.,~\citet{barlow1995} and~Eq.~6 of~\citet{kranmer2012}). We fit both ON and OFF regions simultaneously\footnote{%
Note that the factors $1/N_k^\mathrm{ON}!$ and $1/N_k^\mathrm{OFF}!$ have been omitted here because they drop out in the test statistics Eq.~\ref{eq.lr}.
}: 
\begin{eqnarray}
 \mathcal{L}(\lambda_n;\bonu|\textit{\textbf X}) &=& \mathcal{L}(\lambda_n;f,\alpha,\phip,\sigmap|\{\{\epr_i,\phipr_i\}_{i=0}^{N_k}\}_{k=0}^{N_s})  \\[0.3cm]
{}  &=& P(\bonu) \cdot \prod_{k=0}^{N_s} \exp\left(-g_k(\lambda_n;\bonu)- b_k \cdot \frac{1 + \tau }{\tau}\right) \cdot \prod_{m=0}^{N_k^\mathrm{OFF}} b_k \cdot \nonumber\\
{}   &&  \cdot \prod_{i=0}^{N_k^\mathrm{ON}} \; \left(g_k(\lambda_n;\bonu)+b_k/\tau\right) \cdot \mathcal{P}_k(\epr_i,\phipr_i|\lambda_n;\bonu) ~, \label{eq.L} 
\end{eqnarray} \noindent
where $g_k$ and $N_k$ are the expected and observed number of pulsar events of observation period $k$, respectively, with reconstructed energy within a chosen range $[\epr_\mathrm{min},\epr_\mathrm{max}]$ (see Section~\ref{sec.results})
and reconstructed phase within the ON-phase range $[\phipr_\mathrm{min},\phipr_\mathrm{max}]$. 
Note the dependency of $g_k$ on both the LIV parameter $\lambda_n$ and all nuisance parameters, which are a direct consequence of the limited observed phase range. 
Similarly, $b_k$ which are nuisance parameters, are the corresponding numbers of background events in the standard background control phase (OFF) region $[0.52,0.87]$~\citep{fierro1998}, 
while $\tau$ is the ratio of phase width of the OFF, divided by that of the ON region. The phase limits of the ON region have been optimized 
using simulations (see Section~\ref{sec.checks}). The first product runs over $N_s$ used observation periods, the second and third over the $N_k^\mathrm{OFF}$ events found in the OFF region, 
and $N_k^\mathrm{ON}$ events found in the ON regions, respectively, for each observation period $k$. Here, $P(\bonu)$ is a possible PDF of the nuisance parameters, obtained from external measurements.

A minimum set of nuisance parameters then includes: the \ptwo flux normalization $f$, its spectral index $\alpha$ (see also Eq.~\ref{eq.flux}), 
the mean pulse position $\phip$ and its width $\sigmap$ (see Eq.~\ref{eq:sympulse}), and the $b_k$ background levels. 
Nuisance parameters may nevertheless also include additional asymmetry parameters, a spectral cutoff or other variables parameterizing a different pulse model 
 (see later Sections~\ref{sec.results} or~\ref{sec.uncertainties}). 

The probability density function (PDF) of event $i$ is a normalized combination of PDFs for its measured quantities $(E_i^{\prime},\phi_i^\prime)$ to belong either to a pulsar event $S_k(\epr_i,\phipr_i|\lambda_n;\bonu)$ or a background event $h_k(\epr_i)$~\citep[see, e.g.,][]{Segue2014}: 

\begin{eqnarray}
  \mathcal{P}_k(\epr_i,\phipr_i|\lambda_n;\bonu) &=& \frac{b_k/\tau \cdot h_k(\epr_i)~\, + ~\, g_k(\lambda_n;\bonu) \cdot S_k(\epr_i,\phipr_i|\lambda_n;\bonu)}{g_k(\lambda_n;\bonu)~\, + ~\, b_k/\tau} \quad, \label{eq.pdf}
\end{eqnarray}\noindent
with $h_k(\epr)$ being the (interpolated) spectral energy distribution of the background, and $S_k(\epr_i,\phipr_i|\lambda_n;\bonu)$ the PDF of the pulsar signal 
for the $k$th data subsample, respectively. 
Here, $h_k(\epr_i)$ 
is a complex combination of cosmic-ray events and gamma-ray images from the Crab Nebula, and is very difficult to model analytically.  
Because the integral number of background events is always at least a factor of 20 larger than the integrated signal, $(b_k/\tau)/g_k > 20$, 
an accurate construction of $h_k(\epr_i)$ is however indispensable. 
We chose to linearly interpolate the binned spectral energy distribution of the background region in double-logarithmic space, and interpolate events without any background events using a linear fit to that distribution. 
We find that $h_k(\epr_i)$  follows only approximately a power law, showing subtle features such as spectral breaks. We tested different binnings to the original background distribution and found that their effect is acceptable, but nonetheless a non-negligible source of systematic uncertainties (see Section~\ref{sec.uncertainties}). \par 
The signal PDF, $S_k(\epr_i,\phipr_i|\lambda_n;\bonu)$, is calculated as follows:  
\begin{eqnarray}
S_k(\epr_i,\phipr_i|\lambda_n;\bonu) &=& \frac{\Delta t_k \int_0^\infty \!\!  R_k(E|\epr_i)\cdot \Gamma_{\ptwo}(E,f,\alpha) \cdot F_{\ptwo}(\phipr_i,E|\lambda_n;\phip,\sigmap) \; \diff E}{g_k(\lambda_n;\bonu)} ~,\noindent
\end{eqnarray}
where:
\begin{enumerate}
    \item $\Delta t_k$ is the effective observation time for the $k$th data subsample.
    \item $R_k$ is the telescope response function of the true photon energy $E$ for the $k$th subsample, computed as
	the product of the effective collection area 
and the energy redistribution function of the instrument. 
Both have been obtained from Monte-Carlo simulations and fitted to obtain smooth functions in energy.
	\item $\Gamma_{\ptwo}$ is the pulsar spectrum at \ptwo, namely:
\begin{equation}
\Gamma_{\ptwo}(E)
= f \cdot \big(E/E_\mathrm{dec}\big)^{-\alpha} \cdot \exp(-E/E_b)\quad\mathrm{TeV}^{-1}\,\mathrm{cm}^{-2}\,\mathrm{s}^{-1}\quad.  \label{eq.flux}
\end{equation}
A previous publication using the same data set~\citep{crabtev} obtained the values $f_0 = (5.7\pm0.6)\times 10^{-10}~\mathrm{TeV}^{-1}\,\mathrm{cm}^{-2}\,\mathrm{s}^{-1}$, 
a de-correlation energy~\citep{Fermidecorr2010} $E_\mathrm{dec} = 50\,\mathrm{GeV}$ and $\alpha = (3.0\pm 0.1)$ in a joint fit with \textit{Fermi} data, using a pure power law (i.e. $E_b := \infty$).
A possible exponential cutoff has only been excluded below 700~GeV so far~\citep{crabtev}. 
	\item $F_\ptwo$ is the pulsar phaseogram model for a given LIV intensity $\lambda_n$ and is computed as:
\begin{eqnarray}
F_\ptwo(\phipr_i,E|\lambda_n;\phip,\sigmap) \!
                  &=& \!\int_0^\infty \!\!\!\! \frac{1}{2\pi\sigma_\mathrm{res}\sigma^\prime_\ptwo}
                    \!\cdot\!
                    \exp\!\bigg[\!- \!\frac{\Big(\phi_i\! -\! \phip\! -\! \Delta\phi(E|\lambda_n)\Big)^2}{2\,(\sigma^\prime_\ptwo)^2}\! -\! \frac{\Big(\phipr_i\! -\! \phi \Big)^2}{\,2\sigma^2_\mathrm{res}} \bigg]  \diff \phi \nonumber\\
                  &=& \frac{1}{\sqrt{2\pi}\sigmap}
                    \cdot
                    \exp\bigg[-\frac{\Big(\phipr_i - \phip - \Delta\phi(E|\lambda_n)\Big)^2}{2\,\sigmap^2} \bigg] \qquad, \label{eq:sympulse}
\end{eqnarray}
where $\sigma^\prime_{P2}$ is the intrinsic pulse width at the pulsar itself, which may in principle depend on energy, 
and $\sigma_\mathrm{res}$ the instrumental phase resolution,  which is dominated by the uncertainties of the pulsar ephemerides, the RMS of the timing noise, and the uncertainties of the barycentric corrections. 
Because the latter contribution is two orders of magnitude smaller than the former~\citep{phd.garrido}, the observed width $\sigmap$ can be considered completely dominated by the intrinsic pulse width. 
Note that the pulse form does not necessarily need to follow a Gaussian, and other, even asymmetric, functions cannot be excluded so far. The effect of different alternative possibilities will be investigated later on (see Section~\ref{sec.uncertainties}).

The mean position of the Gaussian includes a signed phase delay produced by the LIV effect under test, described by $\Delta\phi$ (Eq.~\ref{eq.Deltaphi}).

	\item $g_k$ and $b_k$ are the normalization constants of $S_k$ and $h_k$, which depend on the actual realizations of all nuisance parameters, and on $\lambda_n$, once these PDFs are integrated within the phase window limits $\phipr_\mathrm{min}$ and $\phipr_\mathrm{max}$, and the reconstructed energy limits $\epr_\mathrm{min}$ and $\epr_\mathrm{max}$.

$\phipr_\mathrm{min}$ and $\phipr_\mathrm{max}$ could in principle be chosen to be 0 and 1,
respectively, and the PDF constructed cyclic, however in that case the contributions of \pone and the bridge emission~\citep{bridge14} need to be modeled as well, 
unnecessarily complicating the PDF and adding systematic uncertainties to the results. 
Moreover, it is computationally more efficient to reduce the background as much as possible, by choosing tight windows $\phipr_\mathrm{min}$ and $\phipr_\mathrm{max}$ around $\ptwo$.

       \item The limits for the integration over the true photon energy are \textit{formally} set to zero and infinity, but \textit{physically} need to be
             set to a lower value $E_\mathrm{min}$ above which the emission model Eq.~\ref{eq.flux} is considered valid, e.g. a good choice would be the transition
             from the exponential cutoff to the power law, around 40~GeV. Such a value does not hamper the precision of the overall likelihood, since the minimum
             reconstructed energy $E^\prime$ has been chosen to be 400~GeV, sufficiently far from this value in comparison with the energy resolution of 15\%--20\% (see Appendix~\ref{app.datasamples}).
       \item The choice of the cutoff energy $E_b$ is less obvious: the last significant spectral point, obtained with these data,
             lies at $\sim1.5$~TeV and still fits the power law (Eq.~\ref{eq.flux}), although
             an exponential cutoff can only be excluded below 700~GeV at 95\%~CL~\citep{crabtev}.
             A reasonable, justified choice of $E_b$ above 700~GeV is hence \textit{a priori}
             impossible, its effects on the limits on $\lambda_n$ will be studied in Section~\ref{sec.results}.
\end{enumerate}

The PDF for the nuisance parameters flux ($f$) and spectral index ($\alpha$) is assumed to be normally distributed and un-correlated, 
because it was evaluated at the de-correlation energy $E_\mathrm{dec}$~\citep{crabtev}: 
\begin{equation}
 P(f,\alpha) =  \mathcal{N}(\mu_f ,\sigma_{f}^2) \cdot \mathcal{N}(\mu_\alpha,\sigma_{\alpha}^2)   \quad,
\end{equation}\noindent
with $\mu_f$ and $\mu_\alpha$ being the central fit results for $f_0$ and $\alpha$, and $\sigma_f$ and $\sigma_\alpha$ their statistical uncertainties.



The PDF for the pulse position parameters $\phip$ and $\sigmap$ had to be assumed flat because no previous information is available about their values, except for this very same data set. 

The definition of the likelihood, Eq.~\ref{eq.pdf}, assumes that the phases have been reconstructed with sufficient precision (we assume the systematic uncertainty
in the reconstruction of the phases of the order of $10^{-3}$ in phase), such that any residual uncertainty
between reconstructed phase $\phi_i'$ and true phase $\phi$ can be absorbed in the nuisance parameter $\sigmap$.
Similarly, the change of pulsar period from 33.60~ms in 2007 to 33.69~ms in 2014 has been absorbed in $\sigmap$.  Note that both
effects are statistically independent of the reconstructed photon energy. 




\subsection{Application of the profile likelihood to data}\label{sec.results}

The ML algorithm (Eq.~\ref{eq.lr}) is now applied to the MAGIC Crab Pulsar data set~\citep{crabtev}, using $\epr_\mathrm{min}=100$~GeV and $\epr_\mathrm{min}=400$~GeV. 
Toward even lower energy limits, the background results difficult to model with high accuracy,
because gamma-hadron separation works less and less efficiently, especially for those 
data that were taken with only one telescope. 
Remember that the analysis leading to this data sample has been optimized for high energies.
For the minimization of the profile likelihood, we use the \textit{TMINUIT} class of \textit{ROOT}~\citep{minuit,rootcite,tminuit}, employing the \textit{MIGRAD}, and in case of no success, the \textit{SIMPLEX} 
algorithms.

The obtained values of $\lambda_{1,2}$ at the found minima are close to zero in all cases. 
Table~\ref{tab.MLnuisance} (\textit{pulse evolution model 1}) shows the obtained nuisance parameters at the 
minimum. All values obtained for $\epr_\mathrm{min}=400$~GeV are compatible with the ones presented in~\citet{crabtev}. 
The results for $\epr_\mathrm{min}=100$~GeV are compatible with the numbers presented from 
previous analyses of data from 40 to 400~GeV~\citep{veritascrab11,magicpulsar12}. Interestingly, the 
pulse widths seem to \textit{widen} (by an about 1\,$\sigma$ fluctuation into opposite directions) for data below and above 400~GeV. 
This is unexpected, given that a significant \textit{shrinking} of the pulse width had been observed previously from GeV energies to beyond 100~GeV
 (as well as from MeV to GeV energies)~\citep{magicpulsar12,magiccrab11}.  

\begin{table}[h!]
\centering
\begin{tabular}{cll}
\toprule
Nuisance  &  Result  & Result  \\
 Parameter & ($\epr_\mathrm{min} = 400$~GeV) & ($\epr_\mathrm{min} = 100$~GeV)  \\
\midrule
\textit{Pulse evolution model~1}   &                              & \\
\midrule
$\widehat{f}$  &  {\normalsize $6.3   \pm 0.7 $}    &  {\normalsize $6.2  \pm 0.6 $}   \\[0.01cm] 
{\scriptsize $(\times 10^{-10}~\mathrm{TeV}^{-1}\,\mathrm{cm}^{-2}\,\mathrm{s}^{-1})$}     &   \\[0.01cm]
$\widehat{\alpha}$  & {\normalsize $2.81  \pm 0.07   $}  & {\normalsize $2.95  \pm 0.07  $} \\ 
$\hphip$   & {\normalsize $0.403 \pm 0.003  $}  & {\normalsize $0.401 \pm 0.001 $} \\  
$\hsigmap$ & {\normalsize $0.015 \pm 0.003  $}  & {\normalsize $0.011 \pm 0.002 $} \\  
\midrule
\textit{Pulse evolution model~2}   &                              & \\
\midrule
$\widehat{f}$       & {\normalsize $6.3   \pm 0.7$}   & {\normalsize $5.9   \pm 0.5  $}  \\[0.01cm] 
{\scriptsize $(\times 10^{-10}~\mathrm{TeV}^{-1}\,\mathrm{cm}^{-2}\,\mathrm{s}^{-1})$}          &    \\[0.01cm]
$\widehat{\alpha}$  & {\normalsize $2.81  \pm 0.07$}  & {\normalsize $2.92 \pm 0.07  $} \\ 
$\hphip$   & {\normalsize $0.403 \pm 0.004$} & {\normalsize $0.401 \pm 0.001$} \\ 
$\hsigmap$ & {\normalsize $0.015 \pm 0.003$} & {\normalsize $0.009 \pm 0.002$} \\ 
$\widehat{\diff \sigmap / \diff\!\log(E)}$ & {\normalsize $0.00 \pm 0.01$} & {\normalsize $-0.006 \pm 0.004$} \\
\midrule
\textit{Pulse evolution model~3}   &                              & \\
\midrule
$\widehat{f}$                   &  --   & {\normalsize $5.9   \pm 0.6  $}  \\[0.01cm] 
{\scriptsize $(\times 10^{-10}~\mathrm{TeV}^{-1}\,\mathrm{cm}^{-2}\,\mathrm{s}^{-1})$}          &    \\[0.01cm]
$\widehat{\alpha}$              & --    & {\normalsize $2.95 ~\textit{(fixed)} $} \\ 
$\hphip$               & --    & {\normalsize $0.4005 \pm 0.0011$} \\ 
$\widehat{\Delta\phip}$         & --    & {\normalsize $0.004  \pm 0.003$} \\ 
$\widehat{\sigma}_\mathrm{P2,1}$ & --    & {\normalsize $0.0089 \pm 0.0009$} \\ 
$\widehat{\sigma}_\mathrm{P2,2}$ & --    & {\normalsize $0.015 \pm 0.003$} \\ 
$\widehat{E}_t$ (GeV)           & --    & {\normalsize $285   \pm 32$} \\
\bottomrule
\end{tabular}
\caption{Obtained nuisance parameter values at the minimum for $\lambda_{1,2}$ from the Full Likelihood Method. 
The uncertainties are statistical only and have been obtained from the 
diagonal elements of the covariance matrix, provided by \textit{MINUIT}. 
\textit{Pulse evolution model~1} refers to the original likelihood~Eq.~\ref{eq.lr}, while \textit{pulse evolution models~2} and~\textit{3} use the extensions 
Eq.~\ref{eq.model2} and~\ref{eq.model3}, respectively.\label{tab.MLnuisance}}
\end{table}

For this reason,  
we test a second \textit{pulse evolution model~2}, incorporating a linearly changing pulse width with the logarithm of energy (compare also with Figure~3 of~\citet{magicpulsar12}): 
\begin{equation}
\sigmap = \sigma_\mathrm{P2,0} - \frac{\diff \sigmap}{\diff\!\log(E)} \cdot \log_{10}(E/E_\mathrm{min}) \quad. \label{eq.model2}
\end{equation}
\noindent
Including $\diff \sigmap / \diff\!\log(E)$ into the set of nuisance parameters yields $\diff \sigmap / \diff\!\log(E) = 0.00 \pm 0.01$ above 400~GeV, 
compatible with the assumption of a constant pulse width (see Table~\ref{tab.MLnuisance}, \textit{pulse evolution model~2}). 
For reconstructed energies starting from 100~GeV, however the situation changes and an \textit{increasing} pulse width is marginally favored, namely: 
 $\diff \sigmap / \diff\!\log(E) = (-6 \pm 4)\times 10^{-3}$. 
This finding is in agreement with the results presented in~\citet{crabtev} and~\citet{magicpulsar12}.

\begin{figure}
\centering
  \includegraphics[width=0.485\textwidth]{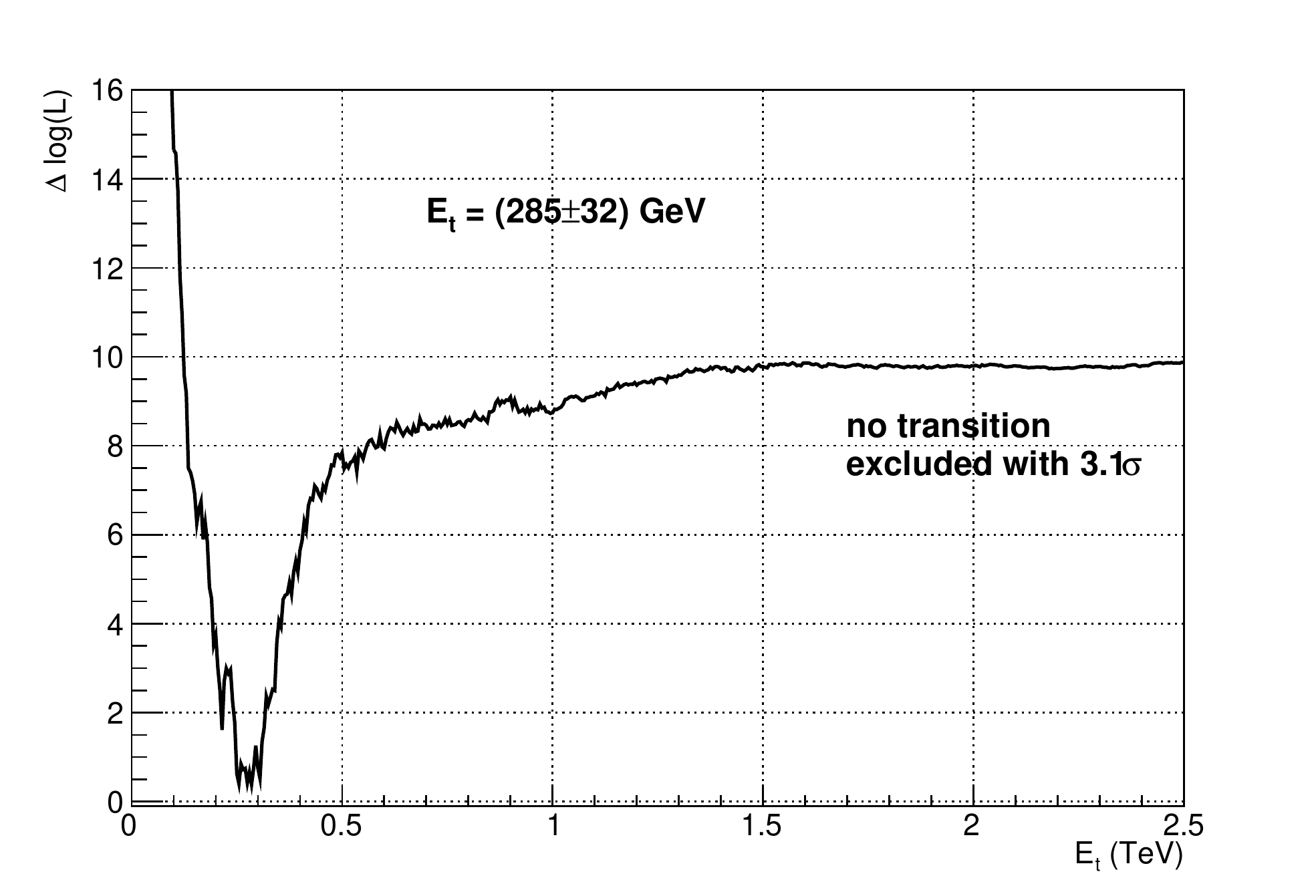}
  \includegraphics[width=0.485\textwidth]{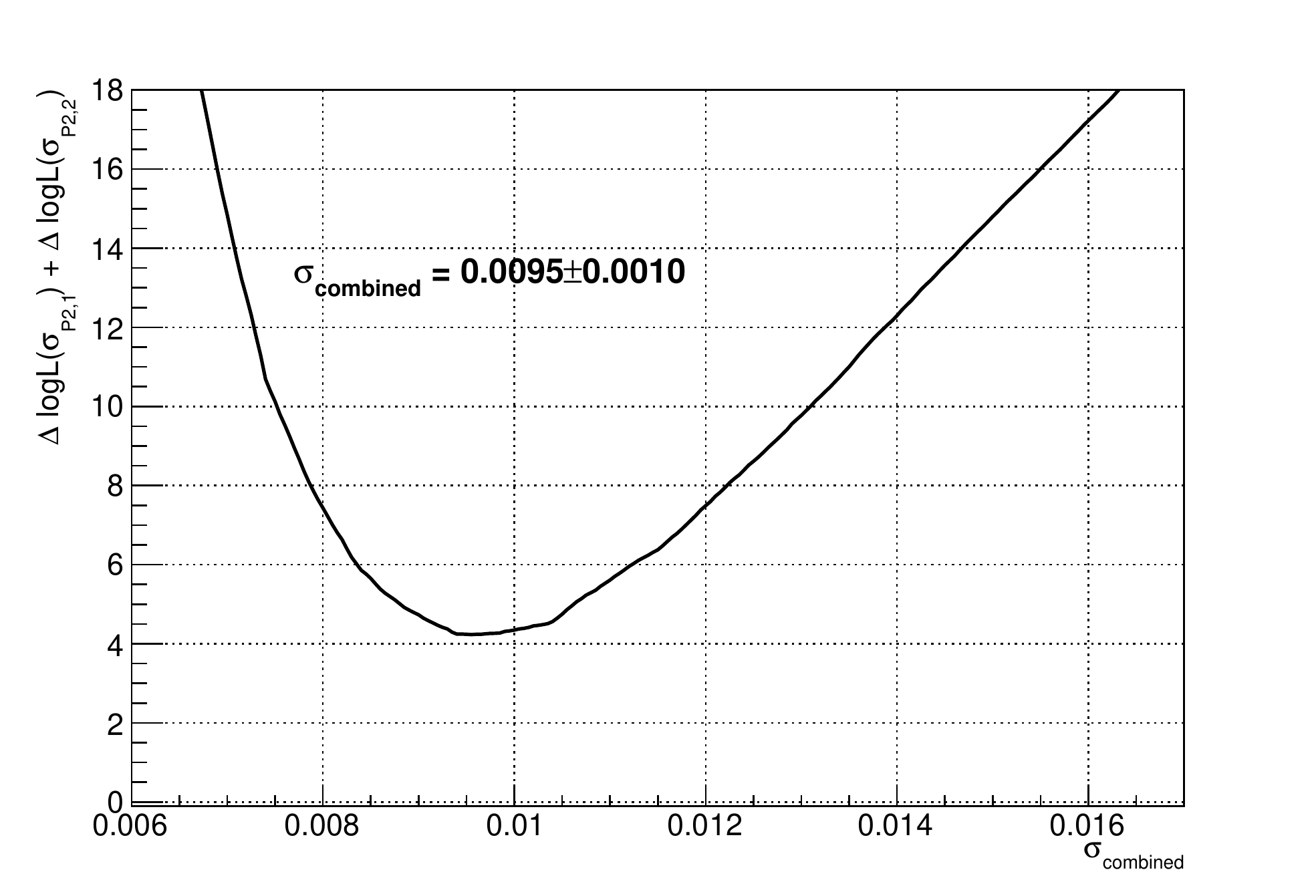}
  \caption{Left: log-Likelihood ratio~(Eq.~\protect\ref{eq.lr}) as a function of the transition parameter $E_t$, under the assumption of no LIV ($\lambda_1=0$).   
Right: the combined log-likelihoods for $\sigma_\mathrm{P2,1}$ and $\sigma_\mathrm{P2,2}$,  with the other nuisance parameters kept fixed at the minimum in both cases.
    \label{fig11}}
\end{figure}

Finally, another \textit{pulse evolution model~3} describing an abrupt transition of both pulse position and pulse width, at a fixed (true) energy $E_t$, can be tested, namely:
\begin{eqnarray}
\sigmap &=& \left\{ \begin{array}{lll}
 \sigma_\mathrm{P2,1} & \qquad\qquad & \textrm{if $E<E_t$}\nonumber\\
 \sigma_\mathrm{P2,2} & \qquad\qquad & \textrm{if $E\ge E_t$}
  \end{array} \right. \\
\phip  &=& \left\{ \begin{array}{lll}
 \phi_\mathrm{P2}     & \quad & ~\textrm{if $E<E_t$}\\
 \phip + \Delta\phip & \quad & ~\textrm{if $E\ge E_t$} \qquad.
  \end{array} \right.   \label{eq.model3}
\end{eqnarray}
\noindent
Just as \textit{pulse evolution model~2} contains \textit{model~1} in the case of $\diff \sigmap / \diff\!\log(E) \rightarrow 0$, 
 \textit{pulse evolution model~3} includes \textit{model~1} in the limit $E_t \rightarrow \infty$. Figure~\ref{fig11}~(left) shows the likelihood at the minimum, when only 
the parameter $E_t$ is varied. One can see that $E_t \rightarrow \infty$ is excluded by about $3\,\sigma$ significance, if the other nuisance parameters are kept fixed, i.e. at the 
values obtained below $E_t$. Similarly, a transition 
of the mean pulse position ($\Delta\phip > 0$) at $E_t = 285$~GeV is found, albeit with only $1.2\,\sigma$ significance. 
Combining the two individual log-likelihoods for $\sigma_\mathrm{P2,1}$ and 
$\sigma_\mathrm{P2,2}$ excludes a common pulse width with about $2\,\sigma$ significance (see Figure~\ref{fig11}~right).

\begin{figure}
\centering
 \includegraphics[width=0.495\textwidth]{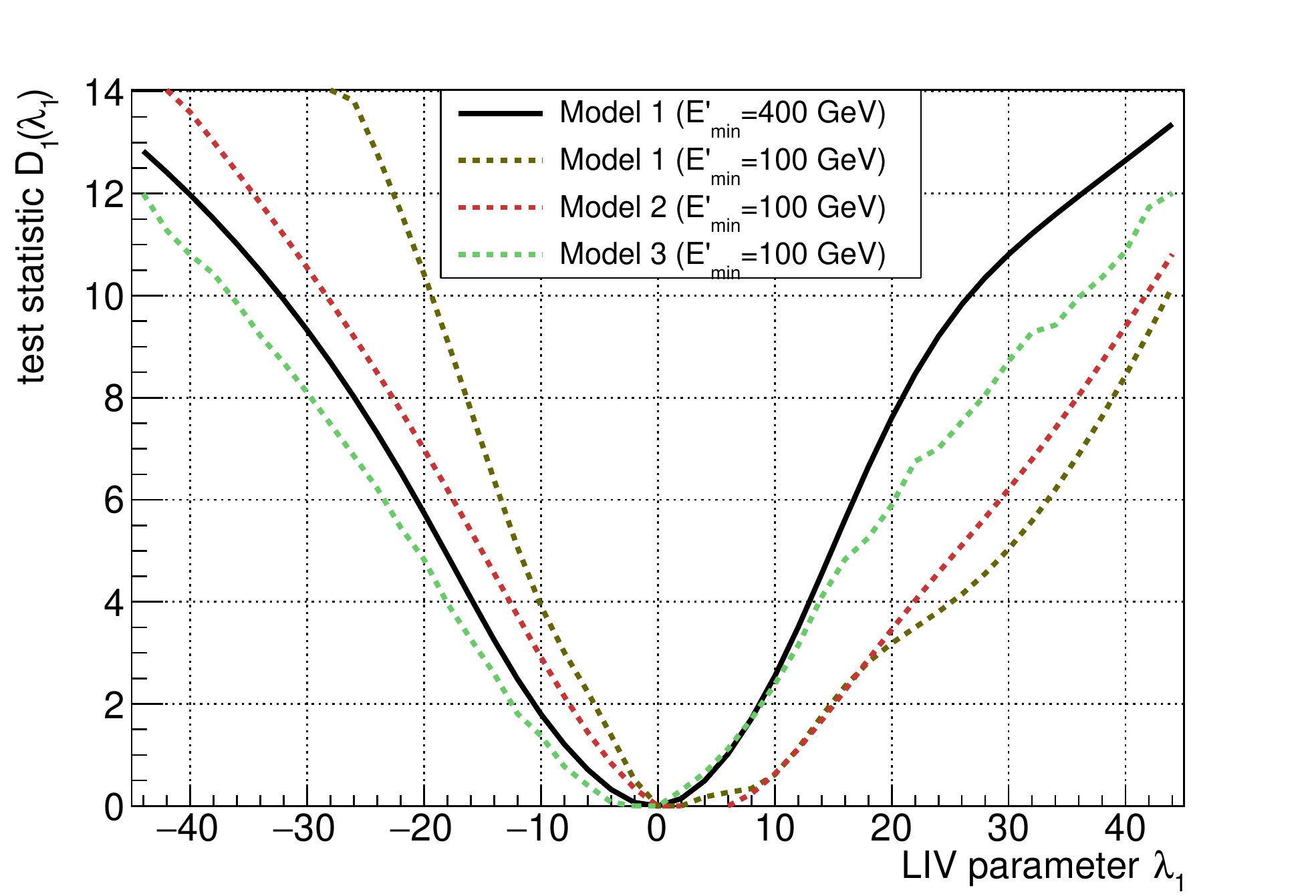}
 \includegraphics[width=0.495\textwidth]{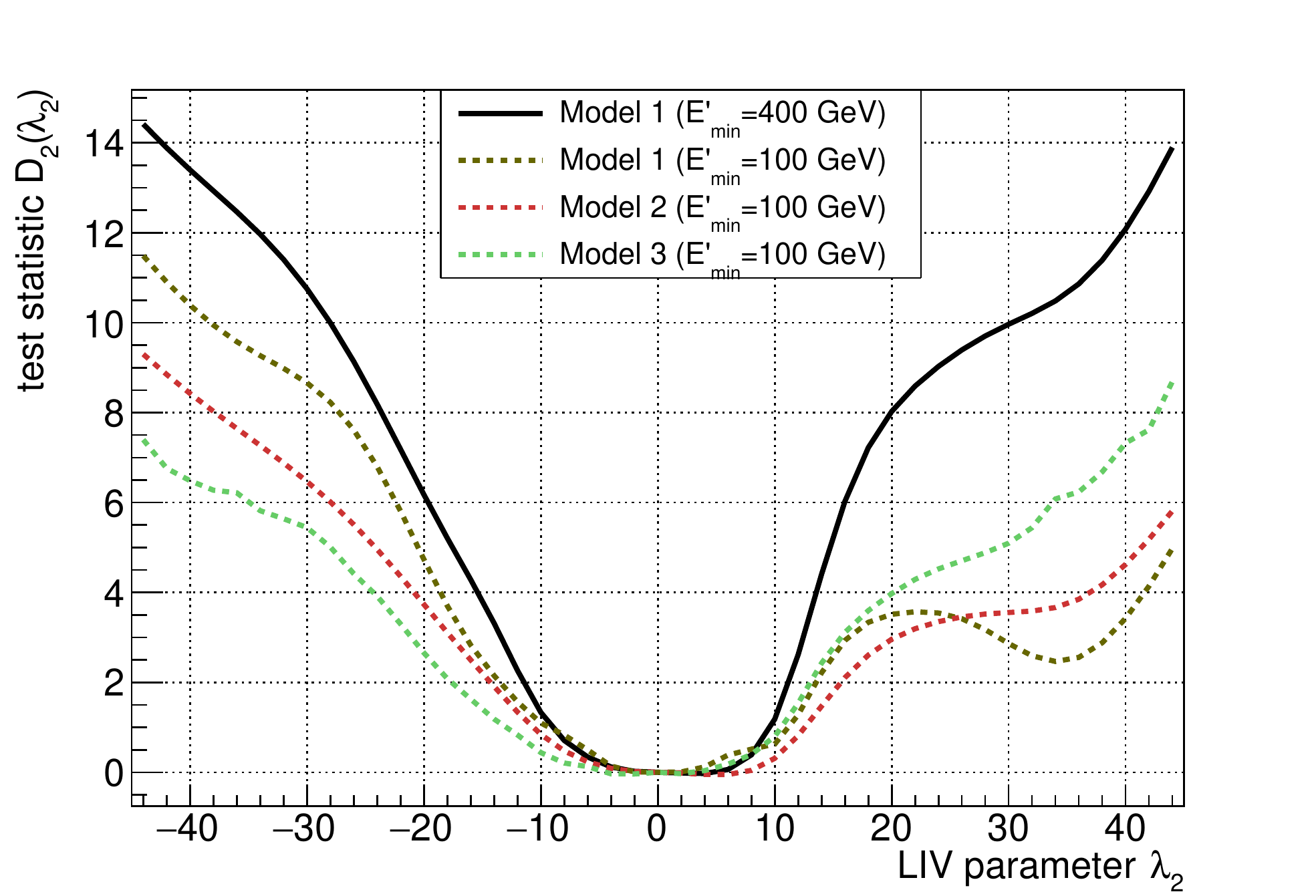}
  \caption{Test statistic (Eq.~\protect\ref{eq.lr}) as a function of the linear LIV parameter $\lambda_1$ (left), and the quadratic LIV parameter $\lambda_2$ (right).
    The different pulse evolution models (\textit{model 1}, \textit{model 2}, and \textit{model 3}) have been used, as have two lower energy limits (see text for details).
    \label{fig7}}
\end{figure}

It is evident that more data are required to clearly determine the behavior of the P2 pulse position and width above about 200~GeV, something out of the scope of this paper.
However, its influence on the behavior of the profile likelihood below 400~GeV is notable, as can be seen in Figure~\ref{fig7}: 
while the profile likelihood  with $E^\prime_{min}=400$~GeV appears symmetric around the minimum, the lower energy limit $E^\prime_{min}=100$~GeV produces a skewed likelihood with non-standard features,
 except for the linear case using \textit{pulse evolution model~3}.  In the quadratic case, a common feature between $\lambda_2 = 20$ and $\lambda_2 = 40$ is observed for all cases, less pronounced for the case of 
\textit{pulse evolution model~3}. 

Strikingly, the incorporation of considerably more data between $E^\prime=100$~GeV and $E^\prime=400$~GeV, which fixes the nuisance parameters $\phip$ and $\sigmap$ to much more 
precise values and should consequently produce a steeper profile likelihood, seems to achieve no improvement -- or even a worsening -- of the precision with which the parameters of interest, $\lambda_{1,2}$ can 
be determined.

Because we cannot be sure about the correct pulse evolution model for P2, at least below about 300~GeV, and to exclude any fake effects on the LIV parameters due to 
wrongly modeled behavior of the nuisance parameters, we decide to restrict the further LIV-search to the part of the sample starting 
with $E^\prime_{min}=400$~GeV. The most probable values are then $\lambda_1 = -0.4$ and $\lambda_2 = -1.5$, both statistically compatible with the null hypothesis
 at the level of
 0.1\,$\sigma$. 

\subsection{Calibration of bias and coverage}\label{sec.checks}

In this section, the statistical properties bias and coverage of the likelihood~Eq.~\ref{eq.lr} are studied with the help of MC simulations.
Using the results from the previous section, we perform sets of 1000 simulations of events lists $\{\epr_i\,,\,\phipr_i,k_i\,|\,\boldsymbol{\nu}\}$. For each simulation set, 
we randomly sample pulse peak positions, pulse widths, absolute flux levels, and spectral indices of normal distributions centered on the values from Table~\ref{tab.MLnuisance}, 
and widths obtained from a Cholesky decomposition of the 
covariance matrix (provided by \textit{MINUIT} from the real data sample), matrix-multiplied with a vector of random normally distributed numbers. 
With this procedure, the correlations between the nuisance parameters, especially between flux, spectral
index, and pulse width, are correctly taken into account \citep[see e.g.][]{walck,blobel}. We also test a $\chi^2$-distributed pulse width, but obtain results very similar results to the normal case. 

To simulate the background, parameterized power laws obtained from the background phase region of real data are used, adapted to each observation period.  
Phases for the background events are picked from a flat distribution of the entire phase range. 
Background is hence simulated simultaneously for the signal and background control phase region. 
The background model is extracted individually from the latter for each simulated data set, 
according to the algorithm described in the previous section. 
Because flux and spectral index can vary quite considerably between each simulation set, the number of reconstructed excess events do so as well, although its mean number coincides 
with the 544 events (for $\epr > 400$~GeV) presented in~\citet{crabtev}.


In order to reduce the computational resources, 
we consider only events up to $\epr = 7$~TeV, 
because 
an extrapolation of the spectrum predicts on average only one event above that energy. In case of an exponential cutoff 
$E_b \lesssim 7$~TeV, the prediction would be even smaller. We can hence get rid of a residual background contribution at high energies, 
to which the likelihood may be sensitive, particularly in the  case of quadratic LIV.


To determine the optimum phase window $\phipr_\mathrm{min},\phipr_\mathrm{max}$, 
we select $\pm 3$ standard deviations around the central fit value obtained in Section~\ref{sec.method1}, i.e. $\phipr \in [0.3558,0.4495]$, 
after explicitly checking that a bigger window does not improve the precision of the method. This effectively occurs
only when the simulated LIV scale is larger than the limits obtained in Section~\ref{sec.method1} (for more details, see~\citet{phd.garrido}).

\begin{figure}[h]
 \centering
 \includegraphics[width=0.49\textwidth]{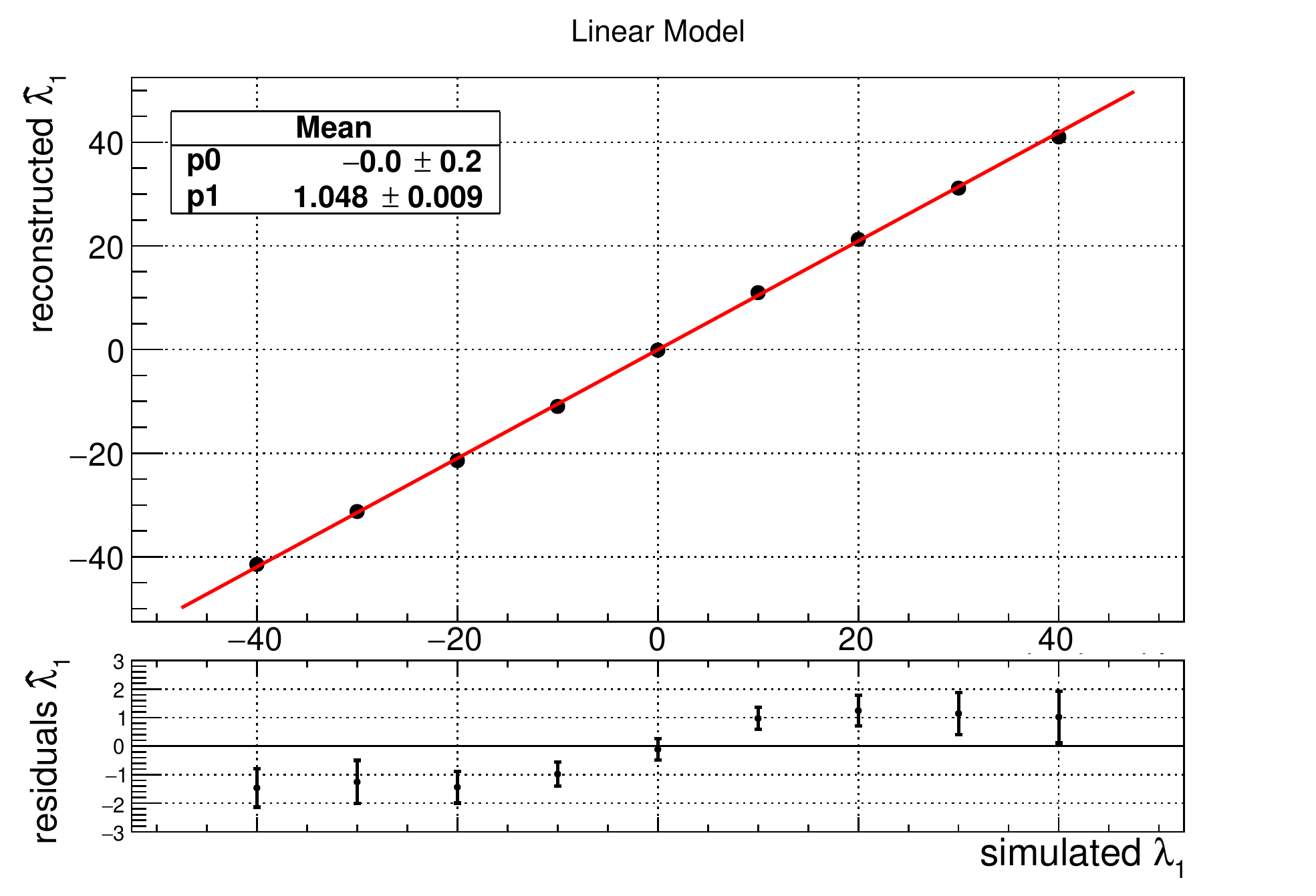} 
 \includegraphics[width=0.49\textwidth]{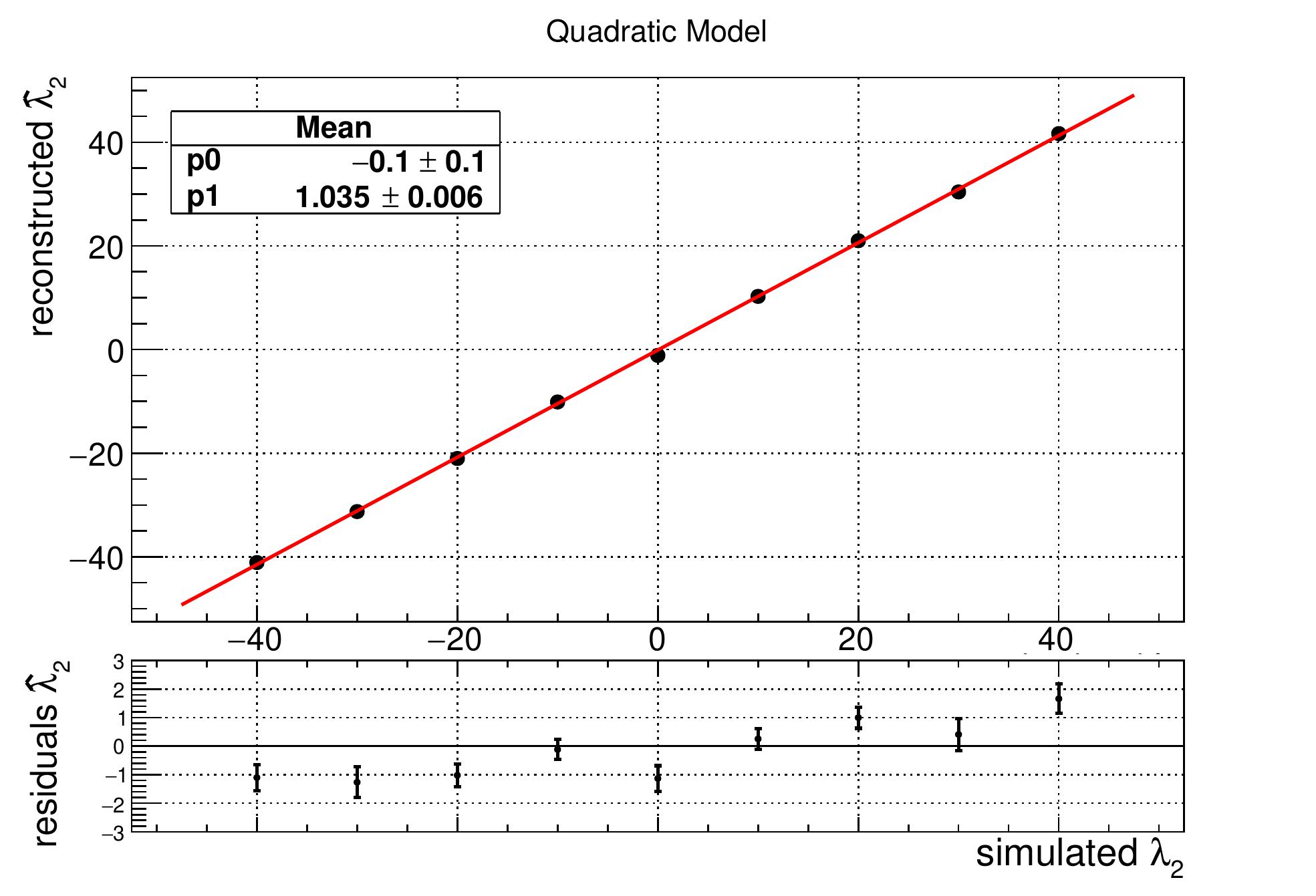} 
\caption{Reconstructed LIV parameter distributions $\widehat{\lambda}$ as a function of
simulated intensities $\lambda$. The mean values have been fitted to a linear response function, 
with the fit results shown in the inserted box. Below the fit, residuals are shown.
}
\label{fig.calibrationcurve}
\end{figure}

Stability tests are carried out with simulated data sets of different LIV parameters, the results of which are shown in 
Figure~\ref{fig.calibrationcurve}. 
We find that our algorithm converges correctly on average, 
under the restriction that 
a very small, but nevertheless significantly measured, bias is present of about 4\%--5\%, \textit{over-estimating}
 the LIV effect.
The bias reduces by approximately half when the median of the distribution of $\widehat{\lambda}$ is evaluated instead of the mean (not shown in Figure~\ref{fig.calibrationcurve}).
We conclude that the ML estimator is consistent, under the  restrictions just mentioned.


In order to demonstrate the statistical behavior of the ML estimator, we simulate and reconstruct the case of no LIV for the following four example cases: 

\begin{figure}[h]
\centering
\includegraphics[width=0.99\textwidth]{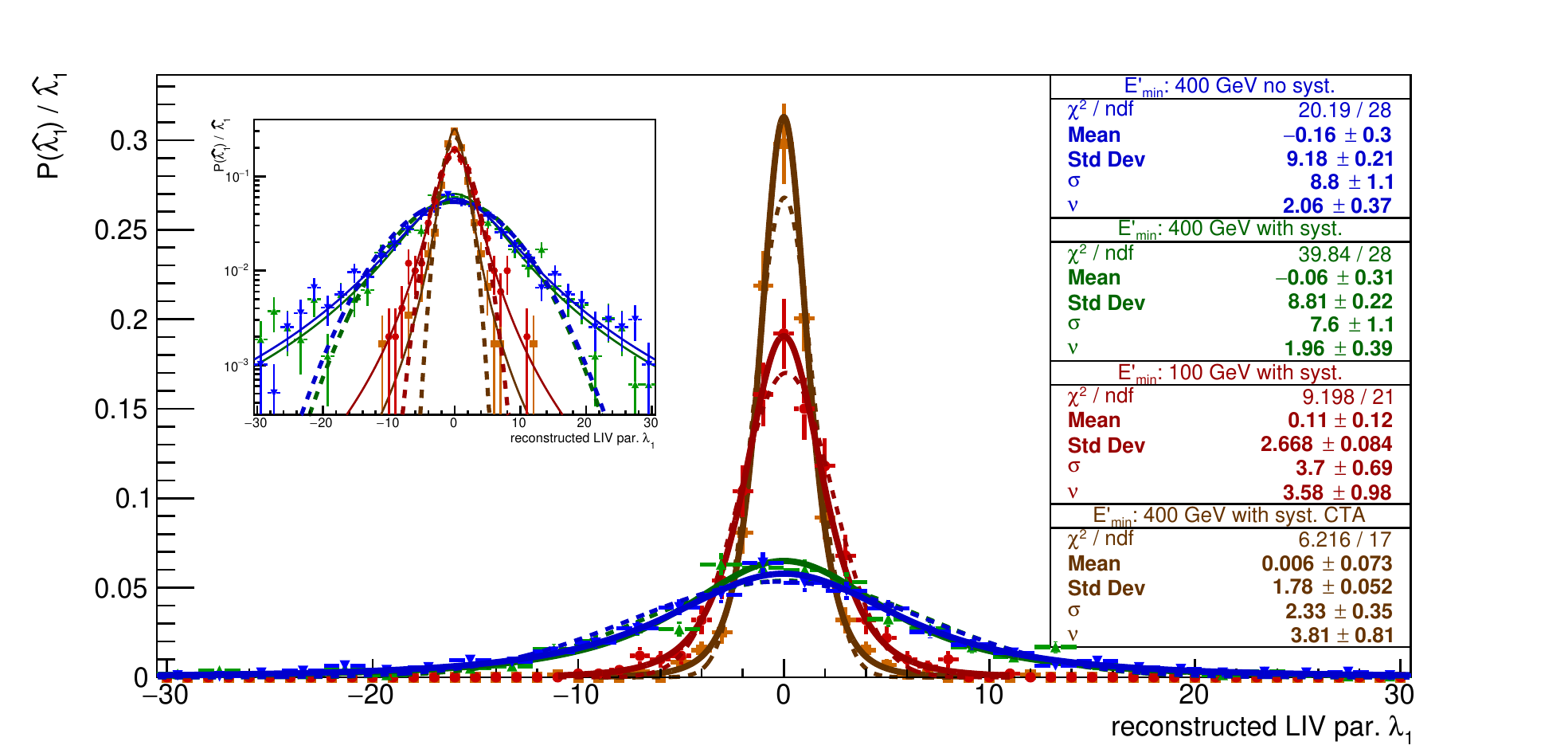} 
\caption{Distribution of estimated linear LIV intensities $\widehat{\lambda}_1$ for 1000 MC simulations with no LIV each, for four example cases: 
$\epr_\mathrm{min}=400$~GeV and no additional systematics simulated (blue); $\epr_\mathrm{min}=400$~GeV and additional systematics simulated on the absolute energy and flux scales (green);
$\epr_\mathrm{min}=100$~GeV with the same systematics simulated (red); and $\epr_\mathrm{min}=400$~GeV, with statistics 10 times higher on signal and background, with energy resolution and systematics simulated for the case of CTA (orange). 
All curves are fit to a normal function (dashed lines) and a Student's $t$-distribution (full lines). All fit results are shown in the Table~\protect\ref{tab.chisqs}. The inlet shows the same figure in logarithmic scale, to make the tails of the distributions  better visible. \label{fig.distlambda1}}
\end{figure}

\begin{description}
\item[$\boldsymbol{\epr_\mathrm{min}=400}$~GeV no systematics:\xspace] samples with $\epr \in [400,7000]$~GeV and effective areas and energy resolution simulated with the same values as those used in the ML analysis.
\item[$\boldsymbol{\epr_\mathrm{min}=400}$~GeV with systematics:\xspace] samples with $\epr \in [400,7000]$~GeV and effective areas and energy resolution varied randomly according to the systematic uncertainties stated in~\citet{perform12,perform16}.
\item[$\boldsymbol{\epr_\mathrm{min}=100}$~GeV with systematics:\xspace]  samples with $\epr \in [100,7000]$~GeV and effective areas and energy resolution varied randomly according to the systematic uncertainties stated in~\citet{perform12,perform16}.
\item[$\boldsymbol{\epr_\mathrm{min}=400}$~GeV with systematics CTA:\xspace]  samples with $\epr \in [400,22000]$~GeV and effective areas and background multiplied by a factor ten, in order to simulate a toy performance of the Cherenkov Telescope Array~\citep{cta}. 
 The energy resolution is estimated from~\citet{ctamc} and systematic uncertainties according to the CTA calibration requirements~\citep{gaugspie}.
\end{description}
\noindent

The last case is included to obtain a toy estimate of the sensitivity of this method within a realistic experimental option in the near future for the same amount of observation time, and if pulse evolution between 200 and 400~GeV remains insufficiently understood.

Figure~\ref{fig.distlambda1} shows the obtained distributions of $\widehat{\lambda}_1$ for these simulated scenarios. 
In all cases, Gaussian fits yield values of $\chi^2/\mathrm{NDF}$ considerably worse than  
\textit{Student's} $t$-distributions with argument $t = \widehat{\lambda}_1/\sigma\cdot \sqrt{\nu}$, which seem to correctly describe the shape of the distribution, and 
particularly the tails. The addition of systematic uncertainties seem to have an effect on the shape, expressed in \textit{Student's} $\nu$-parameter, rather than the distribution 
width~$\sigma$.
The apparent discrepancy can be understood as an effect of the variations of the two nuisance parameters $\sigmap$ and $\alpha$ (see Appendix~\ref{sec:spread}). 
%
%
The slightly non-Gaussian behavior of the reconstructed LIV scales has a direct consequence on the expected coverage properties of the test statistics~Eq.~\ref{eq.lr}. A cumulative 
of the \textit{Student's} distribution with an effective $\nu = (2.06\pm 0.37)/(1.96 \pm 0.39)$ predicts coverage of only ($93\pm 2)$\%/$(92.5 \pm 2)$\% 
at $1.64\,\sigma$ for a one-sided distribution, instead of the 95\% for the 
Gaussian case. We check this behavior using pull-plots and find coverages of only 
$(93.1\pm 0.6)$\%/$(91.2 \pm 0.7)$\%, 
respectively, for the cases of no systematics and included systematics.
Confidence limits using the normal values of $1.64\,\sigma$ (i.e. $\Delta \tilde{D}_n = 2.71$), will hence be \textit{under-covered}. 
%
%
In order to retrieve 95\% coverage using a \textit{Student's} distribution, 
values of $\delta = 1.99\,\sigma$ and $\delta = 2.11\,\sigma$ ($\Delta D_1 = 3.96$ and $\Delta D_1 = 4.45$) are predicted instead.
A sample with limits extracted using these higher values allow us to retrieve the correct coverage of 95\%. 

\begin{figure}[h]
\centering
\includegraphics[width=0.99\textwidth]{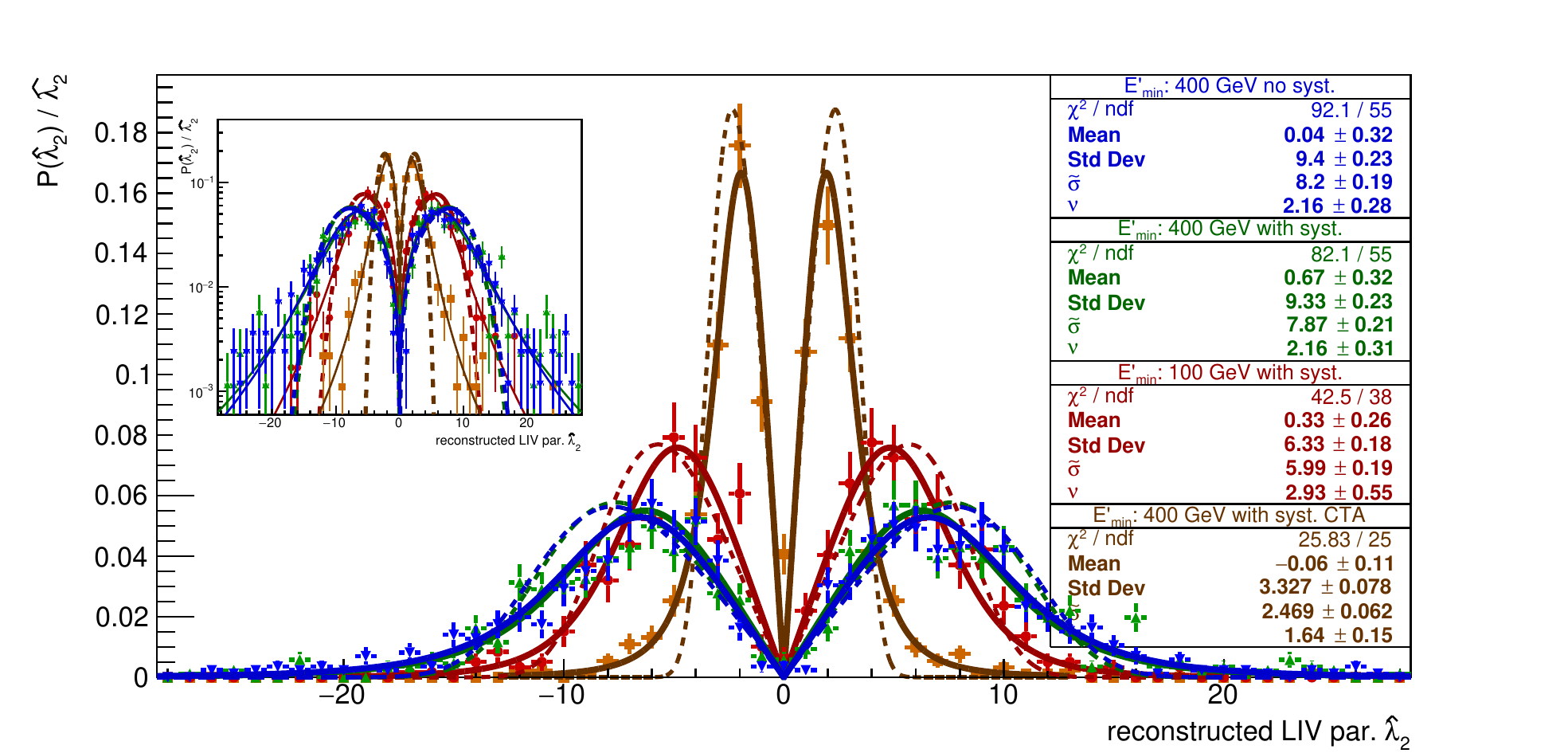} 
\caption{Distribution of estimated quadratic LIV intensities $\widehat{\lambda}_2$ for 1000 MC simulations with no LIV each, for four exemplary cases: 
$\epr_\mathrm{min}=400$~GeV and no additional systematics simulated (blue), $\epr_\mathrm{min}=400$~GeV and additional systematics simulated on the absolute energy and flux scales (green), 
$\epr_\mathrm{min}=400$~GeV, $\epr_\mathrm{min}=150$~GeV with the same systematics simulated, and a ten times higher statistics on signal and background, with systematics on the absolute energy and scale simulated for the case of CTA (orange). 
All curves are fit to a converted normal function (Eq.~\protect\ref{eq:Psimp}, dashed lines) and a converted \textit{Student's} $t$-distribution (Eq.~\protect\ref{eq:Tsimp}, full lines). The fit results of the latter are shown in the text boxes. \label{fig.distlambda2}}
\end{figure}

The distributions of $\widehat{\lambda}_2$ for the simulated scenarios with no LIV are shown in Figure~\ref{fig.distlambda2}.  
In order to better understand  their obviously non-Gaussian shapes, 
it is instructive to study a simplified version of the likelihood Eq.~\ref{eq.L}. In the hypothetical case of just one observation period, 
no background, an infinite energy resolution, and infinite phase range, the likelihood can be written as (see also section~3B of \citet{fermiliv13}): 
\begin{equation}
\mathcal{L_\mathrm{simp}} = (\lambda_n|\boldsymbol{E}^\prime,\boldsymbol{\phi}^\prime) = 
    \prod_{i=0}^{N_\mathrm{events}} \frac{\Delta t_\mathrm{tot} \cdot \avg{A_\mathrm{eff}(E_i^\prime)} \cdot \Gamma_{\ptwo}(E_i^\prime) \cdot F_{\ptwo}(\phipr_i,E_i^\prime|\lambda_n)}{g} \quad. 
\label{eq:Lsimp}
\end{equation}

For this simplified likelihood, the expectation value for $\widehat{\lambda}_n$ can be calculated analytically, yielding: 
\begin{equation}
\frac{\partial \ln \mathcal{L_\mathrm{simp}}}{\partial \lambda_n} \stackrel{!}{=}  0  \nonumber \quad \rightarrow \quad    \left.
\begin{array}{ll}
  \widehat{\lambda}_{n} &=  \bigg( |\widehat{\tau}_n|/c_n  \bigg)^{1/n} \cdot \mathrm{sgn}(\widehat{\tau}_n)\quad,\quad\mathrm{with:} \\[0.4cm]
\widehat{\tau}_n    &=  \mathlarger{\frac{ \avg{\phi_i E_i^n} - \! \phi_0 \avg{E_i^n}}{\avg{E_i^{2n}}}} \quad. \\
 \end{array} \right. \\
\end{equation}

If the phases $\phi_i$ are now distributed normally, so will be the values of $\widehat{\tau}_1 \sim \mathcal{N}(\mu_{\widehat{\tau}_1},\sigma_{\widehat{\tau}_1}^2)$, 
but \textit{not}  $\widehat{\lambda}_{2}$ because of the 
square-root involved. Instead, the PDF of $\widehat{\lambda}_{2}$ has the form\footnote{%
following the rule that if $\boldsymbol{X}$ is a random variable that is distributed like $f_X(\boldsymbol{X})$, then 
a variable transformation $y = g(X)$ yields $f_Y(y) = f_x(g^{-1}(y))\cdot \frac{d}{dy}g^{-1}(y)$.  
}: 
\begin{eqnarray}
\mathcal{P}_\mathrm{simp}(\widehat{\lambda}_{2}) &=& 
  \frac{2}{\sqrt{2\pi}\,(\sigma_{\widehat{\tau}_2}/c_2)} \cdot \exp\bigg(-\frac{\big(\widehat{\lambda}_{2} \cdot |\widehat{\lambda}_{2}| - (\mu_{\widehat{\tau}_2}/c_2)\big)^2}{2 (\sigma_{\widehat{\tau}_2}/c_2)^2} \bigg) \cdot |\widehat{\lambda}_{2}|~.   \label{eq:Psimp}
\end{eqnarray}

Alternatively, if $\widehat{\tau}_n$ is distributed according to a \textit{Student's} $t$-distribution, the PDF for $\widehat{\lambda}_2$ yields:
\begin{eqnarray}
\mathcal{P}_\mathrm{stud}(\widehat{\lambda}_{2}) &=& 
\ \frac{2}{\sqrt{\pi}\,(\sigma_{\widehat{\tau}_2}/c_2)} \cdot \frac{\Gamma((\nu+1)/2)}{\Gamma(\nu/2)} \cdot \bigg(1 + \frac{\big(\widehat{\lambda}_{2} \cdot |\widehat{\lambda}_{2}| - (\mu_{\widehat{\tau}_2}/c_2)\big)^2}{(\sigma_{\widehat{\tau}_2}/c_2)^2} \bigg)^{-(\nu+1)/2} \cdot |\widehat{\lambda}_{2}|~.   \label{eq:Tsimp}
\end{eqnarray}

Eq.~\ref{eq:Psimp} is, strictly speaking, the PDF of the square of a normally distributed variable, 
while Eq.~\ref{eq:Tsimp} 
is the PDF of the square of a variable distributed according to \textit{Student's} $t$-distribution.\footnote{%
%
%
%
For large values of $\mu_{\widehat{\tau}_2} / \sigma_{\widehat{\tau}_2}$, both distributions converge to a Gaussian. 
Substituting $ \tilde{\mu}= \sqrt{|\mu_{\widehat{\tau}_2|}  / c_2} \cdot \mathrm{sgn}(\mu_{\widehat{\tau}_2})$ and $\tilde{\sigma} =\sqrt{\sigma_{\widehat{\tau}_2}  / c_2}$, or for the \textit{Student's} case, $\tilde{\sigma} =\sqrt{\sigma_{\widehat{\tau}_2}  / (c_2 \cdot \sqrt{\nu})}$,
the variance of Eq.~\ref{eq:Psimp} can be derived as $V[\widehat{\lambda}_{2,\mathrm{simp}}]=\sqrt{2/\pi}\cdot \tilde{\sigma}^2$, for the case of $\tilde{\mu} = 0$, and 
for the \textit{Student's} case Eq.~\ref{eq:Tsimp} $V[\widehat{\lambda}_{2,\mathrm{stud}}]= 2 / \sqrt{\pi\nu}\cdot \Gamma((\nu+1)/2) / \Gamma(\nu/2) \cdot \nu/(\nu-1) \cdot \tilde{\sigma}^2$, which 
converges against the Gaussian variance for $\nu \rightarrow \infty$}

\begin{table}[h]
\resizebox{\textwidth}{!}{%
\centering
\begin{tabular}{lccccc}
\toprule
Case  &  \multicolumn{2}{c}{Gaussian Fit} & \multicolumn{3}{c}{Student's Fit}  \\
      &  $\sigma$   &  $\chi^2/\mathrm{NDF}$ &  $\sigma$ &  $\nu$ &  $\chi^2/\mathrm{NDF}$  \\
\midrule
\multicolumn{5}{c}{Linear Model}\\
$\epr_\mathrm{min}=400$~GeV no syst.       & $7.16\pm 0.19$ & 2.93  &  $8.8 \pm 1.2 $ & $2.06 \pm 0.37$ & 0.72 \\
$\epr_\mathrm{min}=400$~GeV with syst.     & $6.88\pm 0.20$ & 2.93  &  $7.6 \pm 1.1 $ & $1.96 \pm 0.39$ & 1.42  \\
$\epr_\mathrm{min}=100$~GeV with syst.     & $2.26\pm 0.07$ & 1.10  &  $3.70\pm 0.69$ & $3.58  \pm 0.98$ & 0.44 \\
$\epr_\mathrm{min}=400$~GeV with syst. CTA & $1.40\pm 0.04$ & 1.51  &  $2.33\pm 0.35$ & $3.81 \pm 0.81$ & 0.37 \\
\midrule
\multicolumn{5}{c}{Quadratic Model} \\
$\epr_\mathrm{min}=400$~GeV no syst.       & $9.27\pm 0.18$ & 3.00  &  $8.20 \pm 0.19 $ & $2.16 \pm 0.28$ & 1.67 \\
$\epr_\mathrm{min}=400$~GeV with syst.     & $9.05\pm 0.14$ & 2.52  &  $7.87 \pm 0.21 $ & $2.16 \pm 0.31$ & 0.94 \\
$\epr_\mathrm{min}=100$~GeV with syst.     & $6.79\pm 0.15$ & 1.86  &  $5.99 \pm 0.18$ & $2.93 \pm 0.55$ & 1.12 \\
$\epr_\mathrm{min}=400$~GeV with syst. CTA & $2.77\pm 0.05$ & 5.04  &  $2.47 \pm 0.06$ & $1.65 \pm 0.15$ & 1.02 \\
\bottomrule
\end{tabular}}
\caption{\label{tab.chisqs} Results of the fits to the distributions of reconstructed values of $\widehat{\lambda}_1$ and $\widehat{\lambda}_2$.}
\end{table}

Contrary to the linear case, the distribution Eq.~\ref{eq:Psimp} 
predicts slight \textit{over-coverage}, while the \textit{Student's} 
of 2nd-order predicts \textit{marginal under-coverage} of $(94.5 \pm 1)$\% which is
found back in the data, namely $(94.2 \pm 0.5)$\%. 

In order to retrieve 95\% coverage using a Student's distribution, values of {\bf $\boldsymbol{\delta = 1.66\,\sigma}$ ($\boldsymbol{\Delta D_2 = 2.76}$)} are predicted. 

\begin{figure}[h]
\centering
\includegraphics[width=0.89\textwidth]{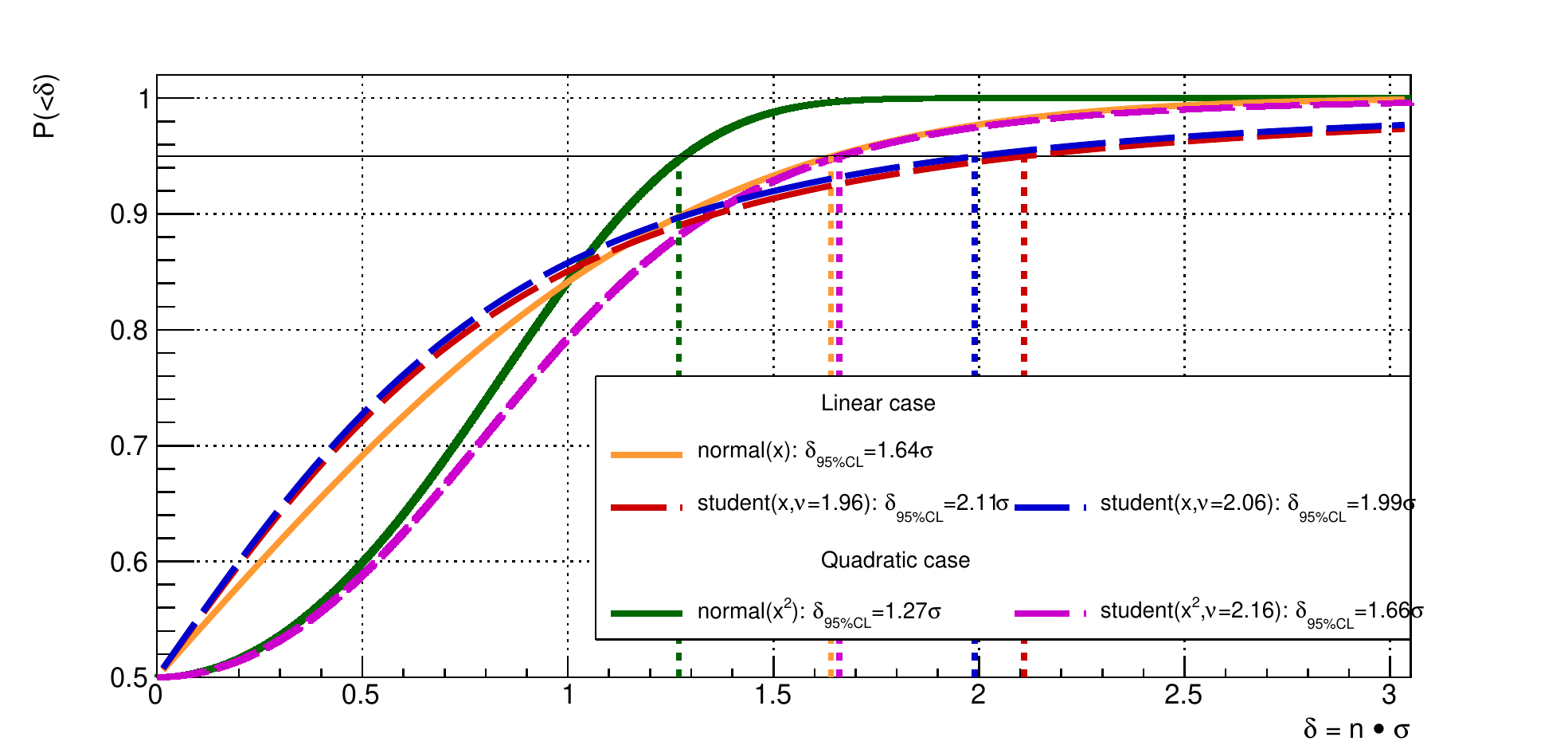} 
\caption{95\% confidence limits for the different used distributions. \label{fig.figure4d}}
\end{figure}

Figure~\ref{fig.figure4d} displays predicted $p$-values and suggested new confidence intervals from different integrated one-sided probability distributions: 
a normal distribution, both of the statistical parameter $x$ as its square $x^2$, and Student's $t$-distributions parameterized with the $\nu$-values obtained from the fits to the 
distributions of reconstructed LIV parameters from the simulations (Figures~\ref{fig.distlambda1} and~\ref{fig.distlambda2}).

\begin{figure}[h]
\centering
\includegraphics[width=0.89\textwidth]{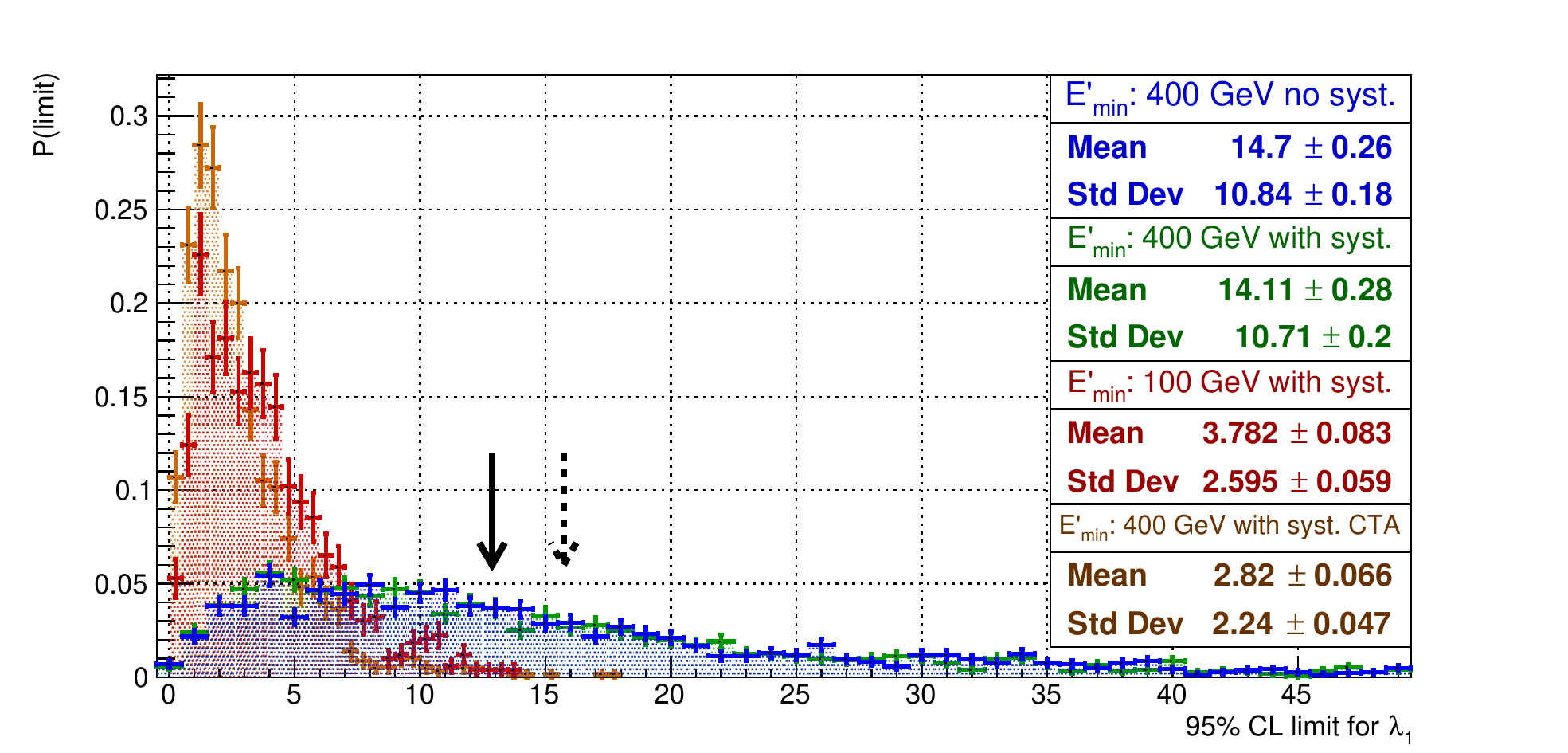} 
\includegraphics[width=0.89\textwidth]{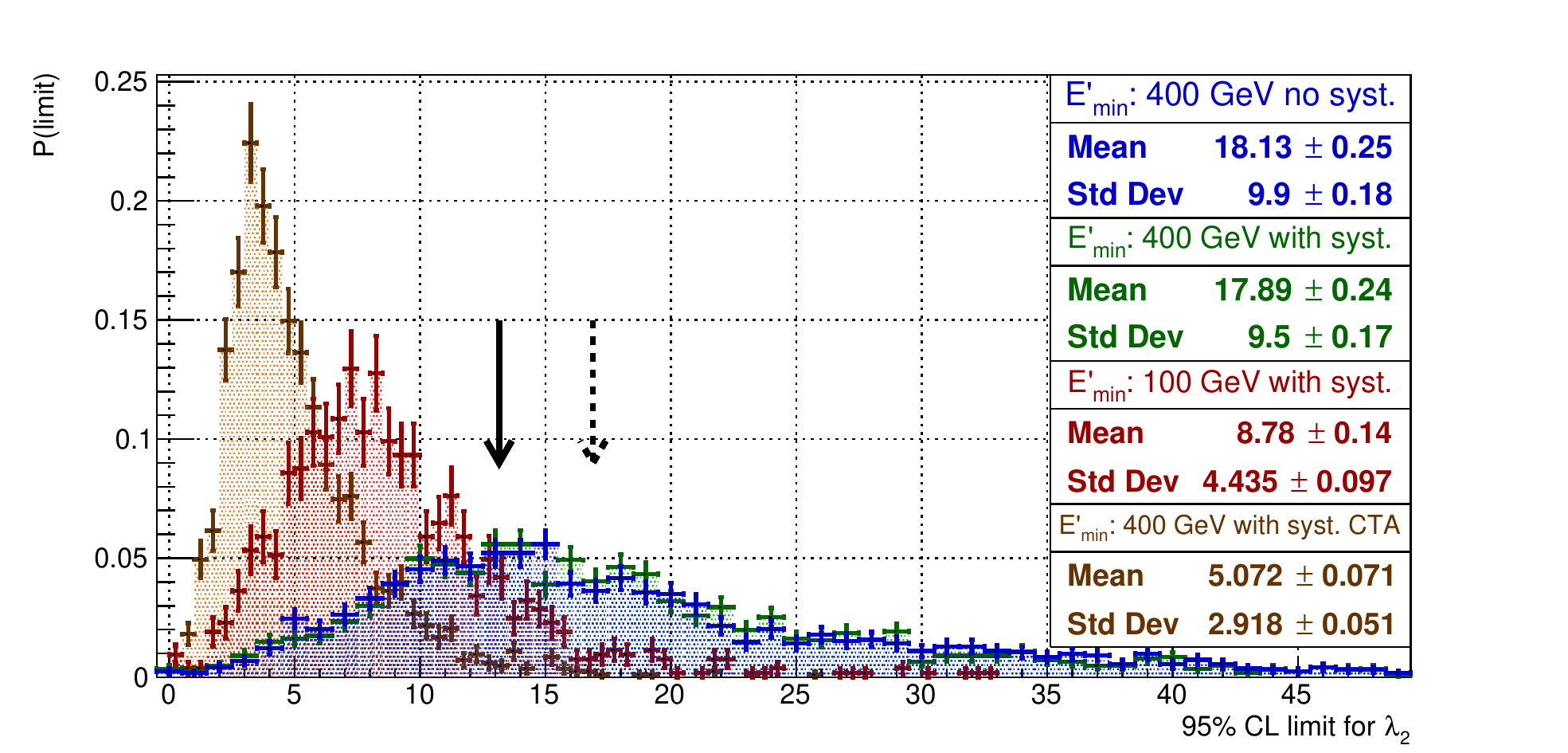} 
\caption{Distribution of 95\% confidence limits for $\lambda_{1}$ (top)  and $\lambda_{2}$ (bottom) for the four simulated case scenarios. The arrows denote the experimentally found limits 
(full line for subluminal behavior, dashed lines for superluminal scenarios).  \label{fig.figure4b}}
\end{figure}

Finally, we calculate limits from each of the 
simulated samples for the previously found test statistic differences $\tilde{D}_n(\lambda_{n}^{95\%CL}) = \Delta\tilde{D}_n$.
Figure~\ref{fig.figure4b} shows the resulting limits. 


\subsection{Systematic uncertainties}\label{sec.uncertainties}




A change of the binning for the spectral energy distribution of the background within reasonable ranges affects the obtained limits 
by $\leq 10$\%. 

As seen in the previous section, inclusion of the systematic uncertainties in 
reconstructed energy and effective area requires evaluating the test statistic at a larger increase, $\Delta D_1=4.45$ instead of $\Delta D_1 = 3.96$, leading to 
a $\sim 6-7$\% increase in the upper limit for the linear case, 
while no effect has been found for the quadratic case. From the obtained statistical precision of the parameters to fit Eq.~\ref{eq:Tsimp}, 
we can estimate that their effect must be smaller than 5\% (95\%~CL) on the limit to $\lambda_2$.

The effect of possible inter-pulse shapes differing from a standard Gaussian is more complicated to assess, because, in principle, a panoply of different 
shapes is possible. Until a better theoretical understanding of the pulse shape is also available, 
we test two easily implemented alternatives: a Lorentzian-shaped pulse, of the form

\begin{equation}
N = (\sigma_\mathrm{L}/\pi) /\left((\phi - \phip)^2 + \sigma_\mathrm{L}^2 \right) ~, 
\label{eq.lorentz}
\end{equation}
and an asymmetric Gaussian shape. 


Evaluating the test statistic on real data using the Lorentzian pulse shape model Eq.~\ref{eq.lorentz} instead of the Gaussian Eq.~\ref{eq:sympulse}, the limits on $\lambda_1$ 
change by maximal 6\%, while those on $\lambda_2$ \textit{improve} by up to 14\%. 


We address the possibility of asymmetric pulse shapes, as usually observed at lower energies~\citep{crabfermi10}, 
using a possible extension of the Gaussian pulse shape with an asymmetry parameter $\Delta \sigma^\prime_\ptwo$: 
\begin{equation}
F^\mathrm{asym}_\ptwo(\phipr_i,E|\lambda_n;\phip,\sigmap) \! = \left\{ 
   \begin{array}{ll}
       \sqrt{\frac{2}{\pi}} \cdot \frac{1}{2\cdot\sigma^\prime_\ptwo + \Delta\sigma^\prime_\ptwo}
                    \cdot
                    \exp\left[-\frac{\Big(\phipr_i - \phip - \Delta\phi(E|\lambda_n)\Big)^2}{2\,(\sigma^\prime_\ptwo)^2} \right] & \\[0.35cm]
                    \qquad\qquad\qquad\qquad\qquad\qquad\qquad\qquad\mathrm{for:}\quad \phipr_i \geq \phip + \Delta\phi(E|\lambda_n) \\[0.4cm]
       \sqrt{\frac{2}{\pi}} \cdot \frac{1}{2\cdot\sigma^\prime_\ptwo + \Delta\sigma^\prime_\ptwo}
                    \cdot
                    \exp\left[-\frac{\Big(\phipr_i - \phip - \Delta\phi(E|\lambda_n)\Big)^2}{2\,(\sigma^\prime_\ptwo+\Delta\sigma^\prime_\ptwo)^2} \right] & \\[0.35cm]
                    \qquad\qquad\qquad\qquad\qquad\qquad\qquad\qquad\mathrm{for:}\quad \phipr_i < \phip + \Delta\phi(E|\lambda_n)~.
   \end{array} \right. \label{eq:asympulse}
\end{equation}

\begin{figure}[h]
 \centering
 \includegraphics[width=0.495\textwidth]{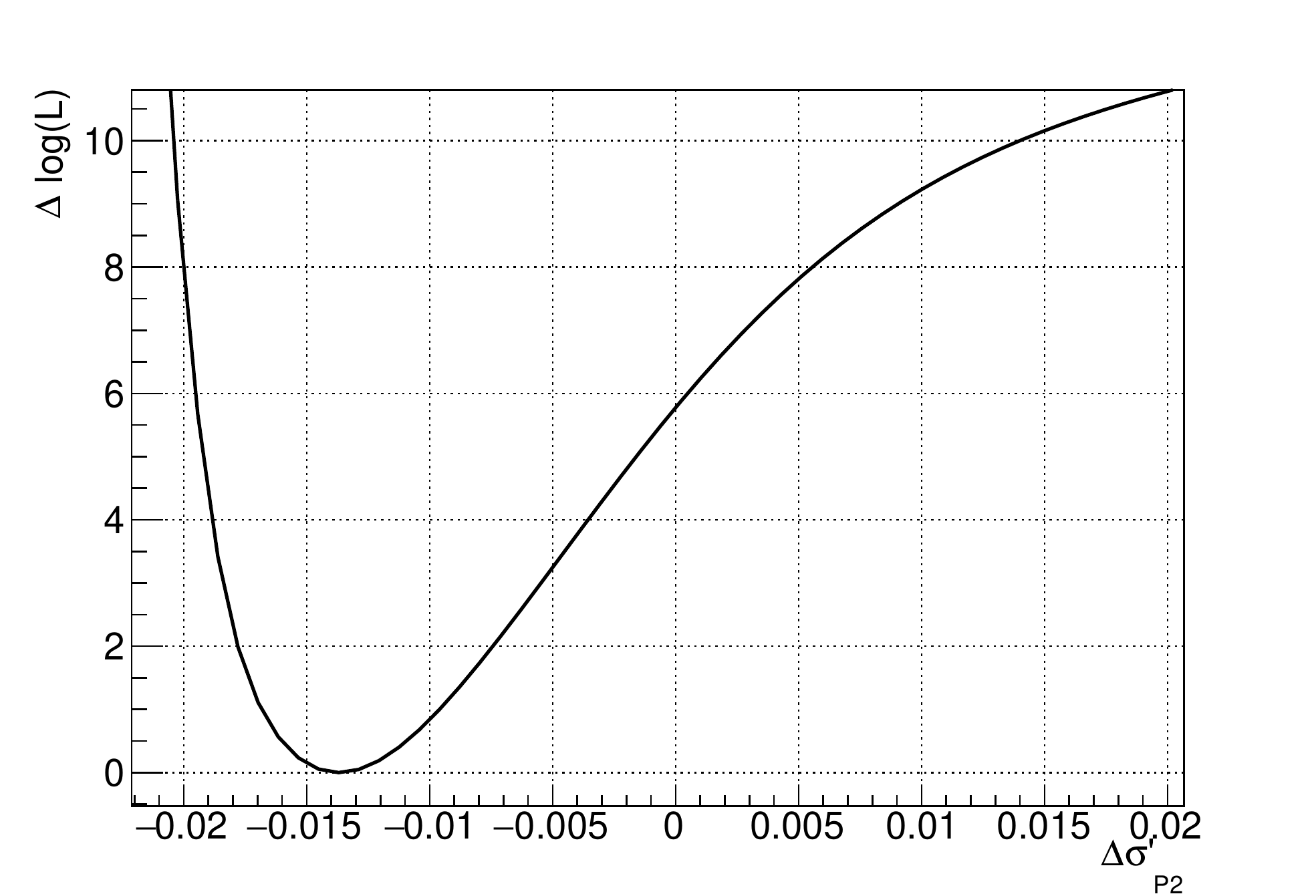} 
\caption{Log-likelihood ratio as a function of the asymmetry parameter $\Delta\sigma^\prime_\ptwo$.
\label{fig:asymlike}}
\end{figure}

A replacement of the pulse shape model $F_\ptwo$ (Eq.~\ref{eq:sympulse}) by $F^\mathrm{asym}_\ptwo$ and addition of $\Delta\sigma^\prime_\ptwo$ to the list of nuisance parameters yields 
$\Delta\sigma^\prime_\ptwo = 0.0137^{+0.0041}_{-0.0032}$, with the hypothesis of a symmetric pulse excluded by $2.4\,\sigma$ (see Fig.~\ref{fig:asymlike}).
The obtained limits on $\lambda_n$ \textit{improve} by at least $\sim30$\% when including the possibility of asymmetric pulse shapes.
However, note that the asymmetry parameter and $\lambda$ are anti-correlated: $\rho(\lambda_1,\Delta\sigma^\prime_\ptwo) = -0.47, \rho(\lambda_2,\Delta\sigma^\prime_\ptwo) = -0.16$,  
and such an improvement is hence expected. However, until a pulse shape asymmetry is however significantly established by more data, we refrain from using it to improve the limits 
on LIV.

An exponential cutoff in the energy spectrum at energy $E_b$ (see Eq.~\ref{eq.flux}) at the currently published constraint of 700~GeV~\citep{crabtev} 
worsens the limits on $\lambda_{1,2}$ by almost a factor of four. We add $E_b$ to the list of nuisance parameters and find
no cutoff as the most probable hypothesis, as stated in~\citet{crabtev}. Assuming no LIV, and profiling the rest of nuisance parameters with respect to 
$E_b$, we obtain a new limit of 4.3~TeV (95\% CL) on the cutoff, and the previous limit of 700~GeV 
disfavored from the most probable hypothesis of no cutoff with a $p$-value of $5\cdot 10^{-4}$. See Appendix~\ref{app.cutoff} for further details.
Evaluating the limits on LIV at $E_b=4.3$~TeV, we observe a worsening 
on the order for 30\% both linear and quadratic LIV. 

Contributions of the bridge emission leaking into the inter-pulse region have not been taken into account in the construction of the likelihood. 
The flux of the bridge from $\phi=(0.026-0.377)$ 
has been measured to $\Gamma_\mathrm{bridge}(E) = (12.2 \pm 3.3) \cdot (E/100\,\mathrm{GeV})^{-(3.35 \pm 0.79)} \times 10^{-11}$ 
TeV$^{-1}$~cm$^{-2}$~s$^{-1}$~\citep{bridge14}.  Any homogeneous coverage of \ptwo by the bridge would have no effect on the LIV analysis, but 
in the worst case, the bridge emission leaks into parts of \ptwo distorting the pulse shape in an energy-dependent way. 
If we conservatively assume 
no spectral cutoff, such a contribution of the bridge to half the \ptwo pulse might amount to 10\% of the observed excess events on average.
We introduce such a case into our toy-MC simulation and obtain a worsening of the limits of less than 5\%. 

The (un-modeled) slow-down of the pulsar frequency, possible glitches of pulsar phase, and mainly the 
uncertainties of the currently available measurements of the distance to the Crab Pulsar all 
contribute to the astrophysical uncertainties.

Table~\ref{tab.syst} summarizes the systematic uncertainties studied. Assuming they are all un-correlated, they add up quadratically to about $\lesssim$42\% for 
the linear and $\lesssim$36\% for the quadratic case.

\begin{table}
\centering
\begin{tabular}{lcc}
\toprule
Systematic Effect  & Size ($\eqgone$) & Size ($\eqgtwo$)  \\
\midrule
Background estimation             & $<$10\%  & $<$15\% \\
Absolute energy and flux scale    & $<$7\%   & $<$5\% \\
Different pulse shapes            & $<$6\%   &  0$^*$  \\
Cutoff in energy spectrum         & $<$30\%  & $<$30\% \\
Contribution from the bridge      & $<$5\%   & $<$5\% \\
Distance Crab Pulsar              & $<$25\%  & $<$12\% \\
\midrule
Total    & $<$42\% &  $<$36\% \\
\bottomrule
\end{tabular}
\caption{List of Studied Systematic Uncertainties. The effects labeled with an asterisk have been found to improve the limits only. \label{tab.syst}}
\end{table}


A different class of systematic uncertainties relates to intrinsic energy-dependent pulse position drifts from the pulsar itself. 
Such intrinsic delays may superimpose a possible LIV effect, and in the worst case, mimic or cancel part of its signature.
We do not include these in Table~\ref{tab.syst} and the derived limits, but will briefly discuss them here. 

Generally, the origin of VHE pulsar emission up to TeV energies is still debated in the literature~\citep{Aharonian:2012jk,Du:2012,Hirotani:2013}, 
albeit it is out of question that inverse Compton scattering must 
be at play in some way or another~\citep{Hirotani:2001,magiccrab11,Lyutikov:2012,Bogovalov:2014}. 
Energy-dependent time drifts of the pulse can then follow those of the illuminating electron population, 
or -- less plausibly because of the Klein-Nishina scattering regime -- the illuminated seed UV and X-ray photons. 
A time dependency of the mean scattering angle is also possible. 

At lower (seed photon) energies, dependencies of the pulse peak positions have previously been studied in detail in the literature~\citep{Mineo:1997,Massaro:2000,Massaro:2006}, particularly throughout the strong X-ray signal regime. 
These studies find constancy of the peak positions of both \pone and \ptwo throughout more than three orders of magnitude, albeit the pulse widths and shapes do change throughout the X-ray domain.\footnote{%
Note, e.g., that the phase boundaries, and particularly $f_3$ remain constant, both in \protect\citet{Massaro:2000} and the more complex multicomponent fits of~\protect\citet{Massaro:2006}.}

To make numerical predictions of the size of a possible LIV-mimicking effect is clearly beyond the scope of this paper. We emphasize, however,
that the measured absence of any linear or quadratic energy dependency of the mean pulse position (up to our sensitivity) 
makes rather unlikely any intrinsic effect of the same size and opposite direction, canceling LIV effects.



\section{Results and discussion}\label{sec.discussion}




%
%

The 95\%~CL limits obtained with the calibrated profiled likelihood method are shown in Table~\ref{tab.MLlimits}, 
with and without including systematic uncertainties. One can see that the profile likelihood method
improves the limits by about a factor of four to five with respect to the simple peak search algorithm (shown in Table~\ref{tab.simplelimits}), 
if the same energy range is used. This is expected because the test statistic Eq.~\ref{eq.lr} exploits additional information, like the 
constancy of the signal over time (through the condition that the expected number of events for the different observation periods is proportional to the respective observation time), 
the characteristics of the fluctuation of the expected signal, and last but not least, the continuous linear (or quadratic) evolution of the signal with energy. 
Along with -- and necessary for -- such an improvement is the stronger confinement of the nuisance parameters, particularly the mean pulse position $\phip$, which strongly correlates 
with the LIV scale (correlation coefficient $\rho \sim 0.5$).

\begin{table}[h!]
\centering
\begin{tabular}{ccccc}
  \toprule
  Case  &  \multicolumn{2}{c}{Crab Pulsar (This Paper)}  &  \multicolumn{2}{c}{GRB090510$^\dagger$} \\
        &   (W/o Systematics) & (Incl. Systematics)  &  (Best of 3 Methods) &  (Likelihood)\\
\midrule
& \multicolumn{3}{c}{$\eqgone$ (GeV)} \\
\midrule
$\xi_1 = +1$  & $ 7.8 \times 10^{17}$ & $5.5 \times 10^{17}$ & $9.3\times 10^{19}$ & $6.3\times 10^{19}$ \\
$\xi_1 = -1$  & $ 6.4 \times 10^{17}$ & $4.5 \times 10^{17}$ & $1.3\times 10^{20}$ & $1.3\times 10^{20}$ \\
\midrule
& \multicolumn{3}{c}{$\eqgtwo$ (GeV)} \\
\midrule
$\xi_2 = +1$  & $8.0 \times 10^{10}$ &  $5.9 \times 10^{10}$ & $1.3 \times 10^{11}$ &  $8.6 \times 10^{10}$  \\
$\xi_2 = -1$  & $7.2 \times 10^{10}$ &  $5.3 \times 10^{10}$ & $9.4 \times 10^{10}$ &  $9.4 \times 10^{10}$  \\
\bottomrule
\end{tabular}
\caption{Ninety-five percent CL limits from the Profile Likelihood Method, together with the best limits from GRB090510 obtained by \protect\citet{fermiliv13}. $^\dagger$10\% system. uncertainty due to instrumental effects not included.\label{tab.MLlimits}}
\end{table}


Our new limits improve previous constraints from the Crab Pulsar~\citep{nepomuk} by almost a factor of three for the linear case, and by about an order of magnitude for the quadratic case
(depending on how systematic uncertainties, not mentioned in \citet{nepomuk},  are accounted). 
For the linear case, our limits are still two orders of magnitude below the best experimental results 
obtained from GRB090510~\citep{fermiliv13} (see Table~\ref{tab.MLlimits}). For the quadratic case, however, the current best constraints from~\citet{fermiliv13}
are only about a factor of two better than our limits.  
It should be noted, however that the limit of $\eqgtwo > 1.3\times 10^{11}$~GeV, reported in their abstract, does not incorporate the additional systematic uncertainty 
of 10\% from instrumental effects, estimated in their Section~6B. Moreover, that limit applies only to the subluminal case (the superluminal limit from the same method is a factor seven 
worse) and came out as the best of three statistical methods applied. 
If we consider their likelihood results only, and include the mentioned systematic uncertainty, our limits are only 
%
30\% and 60\% lower for the subluminal and superluminal cases, respectively.
Even better, though rather model-dependent constraints on subluminal quadratic LIV ($\xi_2 = +1$) have been very recently published 
by \citet{rubstov2017} ($\eqgtwo > 2.1 \times 10^{11}$~GeV) and \citet{martinez2017} ($\eqgtwo > 2.8\times 10^{12}$~GeV), 
using the highest energy photons observed from the Crab Nebula, 
and the apparent absence of a modification of the Bethe-Heitler cross-section for pair production in the atmosphere, 
or the absence of photon decay at these energies.  
A detailed treatment of the complicated systematics inherent to the IACT technique at multi-TeV energies (e.g. due to image leakage out of the camera or saturation of the  
readout), and the source spectra themselves, has, however, not yet been addressed.

Given the strong arguments for an experimental exclusion of any linear LIV effects~\citep{Goetz:2014,Kislat:2017}, limits constraining the quadratic 
dispersion of photons with energy have now become of greater interest. 
Moreover, if LIV is not isotropic, 25 non-birefringent coefficients must be constrained via direction dependent limits~\citep{Kislat:2015}
whose current best values are five to six orders of magnitude worse.

Unlike flaring astrophysical sources like AGNs or GRBs, pulsar data can be continuously accumulated and statistics improved thereby. 
The likelihood is currently still dominated by background fluctuations, as well as un-resolved systematics, particularly the 
pulse shape and its evolution with energy. Such effects have possibly been found between 200 and 300~GeV, although the given statistics does not allow to claim 
firm detection. More data will eventually allow to shed light on the pulse evolution in this energy range and 
subsequently include events with reconstructed energies below 400~GeV into the likelihood analysis. 
Our simulations (see Fig.~\ref{fig.figure4b}) have shown that this possibility alone may already improve the limits by at least a factor of two. 
Moreover, more data will allow to better model the pulse shape itself and take less conservative choices than the used  
 Gaussian pulse shape with fixed symmetric width. 

The possibility to take regular data on the Crab Pulsar with a telescope system at the zenith of its performance~\citep{perform16}
will now permit to regularly improve the sensitivity, 
and even plan such observations based on numerical predictions of such improvements. 
A data set of 2000~hr of stereo data, something perfectly within reach for the MAGIC collaboration, given the regular observation of the Crab Pulsar for calibration purposes, 
can hence ensure an improvement of the quadratic limit by a factor of two, using data above 400~GeV reconstructed energy alone. 
Within a framework of collaboration between the different current IACT installations, a significantly higher amount of data is even plausible.
Such a limit will reach the current world-best constraints, but has the possibility to go well beyond these, because these data can also help to better understand
pulse evolution of the inter-pulse, and such would allow to include events below 400~GeV in the likelihood.

Moreover, it is hoped that the \textit{Gaia} mission~\citep{GaiaMission:2016}
will soon be able to measure the distance to Crab to at least an order of magnitude better precision, removing one of the main uncertainties to these limits.



\section{Summary and conclusions}\label{sec.conclusions}

We have  made use of the profile likelihood method and the Crab Pulsar signal above 400~GeV detected by the MAGIC gamma-ray telescopes~\citep{crabtev} to perform a test on 
LIV involving an additional linear or quadratic dispersion relation term  with energy for photons. 
No significant correlation between arrival time and energy of the pulsar photons is observed, and upper limits on the linear and quadratic energy scale of LIV have been derived. 
The profile likelihood has been carefully calibrated with respect to its bias and coverage properties. For the first time for the Crab Pulsar,
systematic uncertainties have been studied and included in the limits, apart from overall conservative choices in the selection of pulse shape models and tested energy ranges.

While the obtained limits are less constraining for the linear case, 
they come to lie at less than a factor two from the current best limit from GRBs for the interesting quadratic case~\citep{fermiliv13}, depending on which of the several limits in~\citet{fermiliv13} are chosen. 
There is nevertheless a large potential for improvement, once the form of the pulse shapes and their evolution with energy are better understood. 
We observe hints for such an evolution, particularly in the range between 200 and 300~GeV, although statistics do not allow yet to make a firm claim. 

This paper brings back pulsars to the class of astrophysical objects that are useful (and competitive) to investigate time of flight differences of 
energetic photons. Due to the stable and continuous nature of pulsar emission, limits can now be constantly improved over time, and corresponding observations 
planned accordingly. Source intrinsic effects that might mimic a possible LIV signal from flaring sources \citep[see, e.g.,][]{bednarekwagner2008,zheng2011}  
are thus being diversified, adding to the robustness of obtained limits and/or possible future signals. 

A combination of the profile likelihood obtained in this paper, with those from similar searches using other sources and instruments, 
like the strong AGN flares observed by MAGIC and H.E.S.S.~\citep{liv-mkn501,hessliv2011}, 
and combinations of GRBs~\citep{fermiliv13} can be another promising way to further constrain the effects of LIV, particularly with concern to the quadratic energy dependency of the photon time of flight.
The arrival of the next-generation VHE gamma-ray observatory, the Cherenkov Telescope Array (CTA), will easily improve this limit even if a of factor ten less observation time is dedicated to the Crab pulsar.

\acknowledgments

We would like to thank the Instituto de Astrof\'{\i}sica de Canarias for the excellent working conditions at the Observatorio del Roque de los Muchachos in La Palma. The financial support of the German BMBF and MPG, the Italian INFN and INAF, the Swiss National Fund SNF, the ERDF under the Spanish MINECO (FPA2015-69818-P, FPA2012-36668, FPA2015-68378-P, FPA2015-69210-C6-2-R, FPA2015-69210-C6-4-R, FPA2015-69210-C6-6-R, AYA2015-71042-P, AYA2016-76012-C3-1-P, ESP2015-71662-C2-2-P, CSD2009-00064), and the Japanese JSPS and MEXT is gratefully acknowledged. This work was also supported by the Spanish Centro de Excelencia ``Severo Ochoa'' SEV-2012-0234 and SEV-2015-0548, and Unidad de Excelencia ``Mar\'{\i}a de Maeztu'' MDM-2014-0369, by the Croatian Science Foundation (HrZZ) Project 09/176, the University of Rijeka Project 13.12.1.3.02, by the DFG Collaborative Research Centers SFB823/C4 and SFB876/C3, and by the Polish MNiSzW grant 2016/22/M/ST9/00382.


\begin{appendix}

\section{Used data samples}
\label{app.datasamples}

Because the energy reconstruction and effective collection area of the system were different for each combination of camera hardware \citep[see, e.g., Fig.~1 of][]{crabtev}), trigger, and readout system, the data had to be divided in several subsamples, each with similar instrumental response.
This data set was down-selected to more than 300~hr of excellent quality data, including particularly medium and high-zenith angle 
observations that provide better sensitivity above about 800~GeV~\citep{perform16}.

Table~\ref{tab.datasamples} lists the resulting 19 data samples used for this study. The first two sets (I and II) were taken with 
one telescope in stand-alone mode \citep[see][]{Aliu2009293} and had an energy resolution of around 20\% in the energy range from 400~GeV to 1~TeV. 
The rest of the data samples were taken with two telescopes, operated as a stereoscopic system \citep[see][]{perform12}. These have an energy resolution of 15--17\% at $\sim$1~TeV~\citep{perform16}. 
Major upgrades were carried out in mid-2012 and mid-2013, first replacing the readout system, and in the next year, the camera of the first telescope, to achieve a system with almost identical telescopes.
The data samples III--VI were taken before the major upgrade~\citep{perform13}, while samples VII--XIX are from after the upgrade~\citep{perform16}.  
From 2012 on, an upgraded \textit{sumtrigger}~\citep{sumtrigger} was tested on Crab, together with the normal coincidence trigger, and was used to reduce the effective energy threshold of the system~\citep{magicscience}. 
However, those events triggered by the sum trigger, and not the standard coincidence trigger, were not included in this analysis.

Using the ephemeris provided by the Jodrell Bank Observatory~\citep{jodrellbank}, a phase value was assigned to each of the recorded events with an
accuracy of about 4~$\mu$s (see Sect.~4.6 of~\citet{phd.garrido}) using the TEMPO2 package~\citep{hobbs2006} (and cross-checked by our own code~\citep{phd.marcos}).

\begin{table}[h!]
\centering
\begin{tabular}{cccccc}
\toprule
Data Set   & Observation  & Zenith Angle       & Effective    & Telescope & Observation  \\
           & Cycles       &     Range          &   On-time    &  System   & Configuration \\
           &              &     (deg.)         &     (hr)      &           &               \\
\midrule
I         &  2--4         &    5--35           &      31      &   mono    & wobble  \\
II        &  2--4         &    5--35           &      66      &   mono    & on \\
III       &  5--6         &    5--35           &      40      &  stereo   & wobble \\
IV        &  5--6         &    35--50          &      16      &  stereo   & wobble  \\
V         &  5--6         &    50--62          &       5      &  stereo   & wobble \\
VI        &  5--6         &    5--35           &      34      &  stereo   & on   \\
VII       &    7          &    5--35           &       4      &  stereo   & sumtrigger \\
VIII      &    7          &    35--50          &       2      &  stereo   & sumtrigger \\
IX        &    7          &    5--35           &       5      &  stereo   & sumtrigger \\
X         &    7          &    35--50          &       8      &  stereo   & sumtrigger \\
XI        &    8          &    5--35           &      22      &  stereo   & sumtrigger \\
XII       &    8          &    35--50          &       5      &  stereo   & sumtrigger \\
XIII      &    8          &    50--70          &      12      &  stereo   & sumtrigger \\
XIV       &    8          &    5--35           &      22      &  stereo   & sumtrigger \\
XV        &    8          &    35--50          &       5      &  stereo   & sumtrigger \\
XVI       &    8          &    50--70          &       9      &  stereo   & sumtrigger \\
XVII      &    9          &    5--35           &      26      &  stereo   & wobble \\
XVIII     &    9          &    35--50          &       6      &  stereo   & wobble \\
XIX       &    9          &    50--70          &       8      &  stereo   & wobble \\
\bottomrule
\end{tabular}
\caption{Summary of the Used Data Samples: Observation cycles are MAGIC-internal numbers where each cycle corresponds to roughly one year. 
The telescopes were operating first with only one telescope (in \textit{mono}-mode), while later a second telescope was added and 
\textit{stereo} observations made possible. Observations can either be carried out in \textit{wobble} mode~\protect\citep{wobble94}, or 
\textit{on} mode, where the source is imaged on to the center of the camera. 
Data sets labeled \textit{sumtrigger} contained additional events triggered by the sum trigger~\protect\citep{sumtrigger}. \label{tab.datasamples}}
\end{table}

\section{Dependency of the spread of estimated LIV parameters on nuisance parameters}
\label{sec:spread}

\begin{figure}[h!]
\centering
\includegraphics[width=0.49\textwidth]{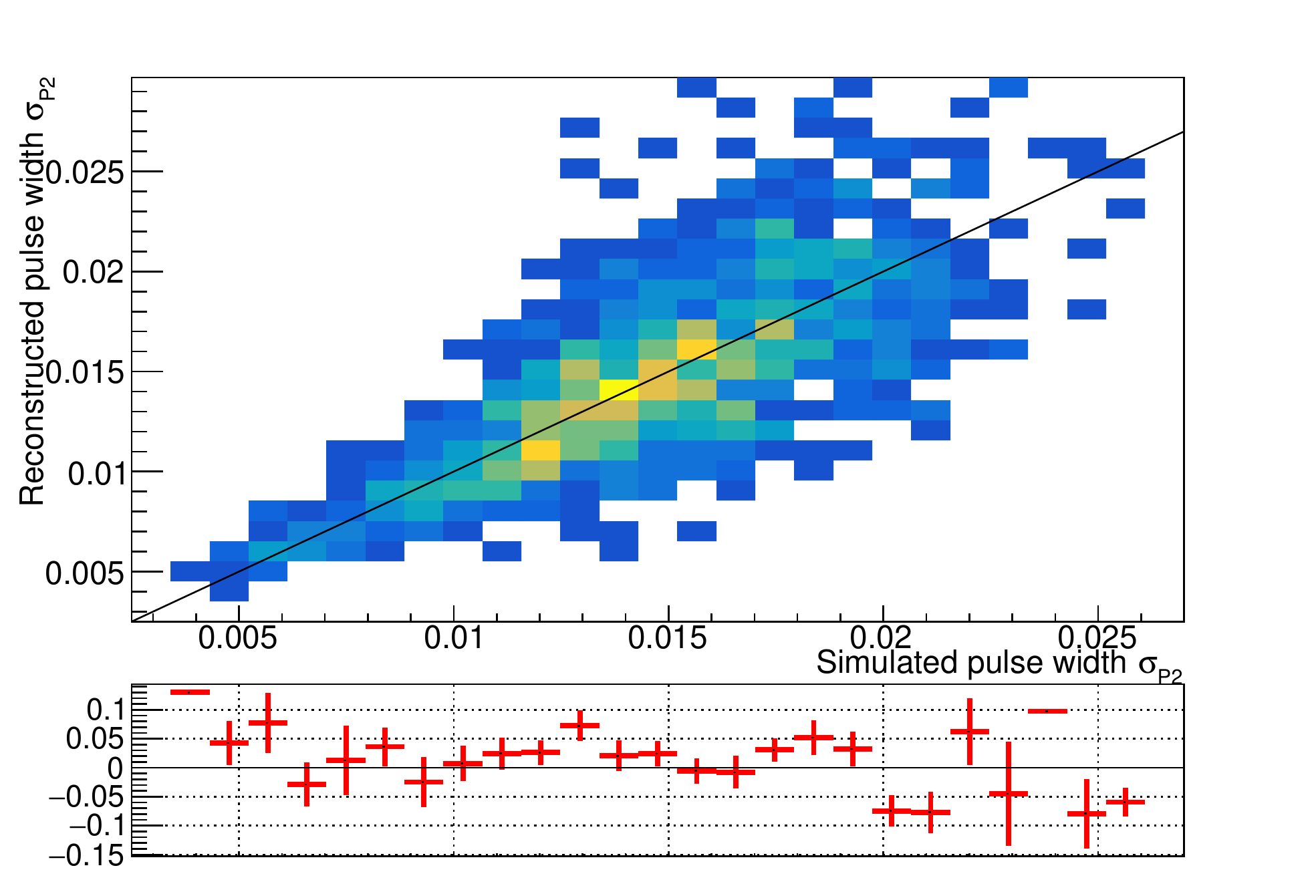} 
\includegraphics[width=0.49\textwidth]{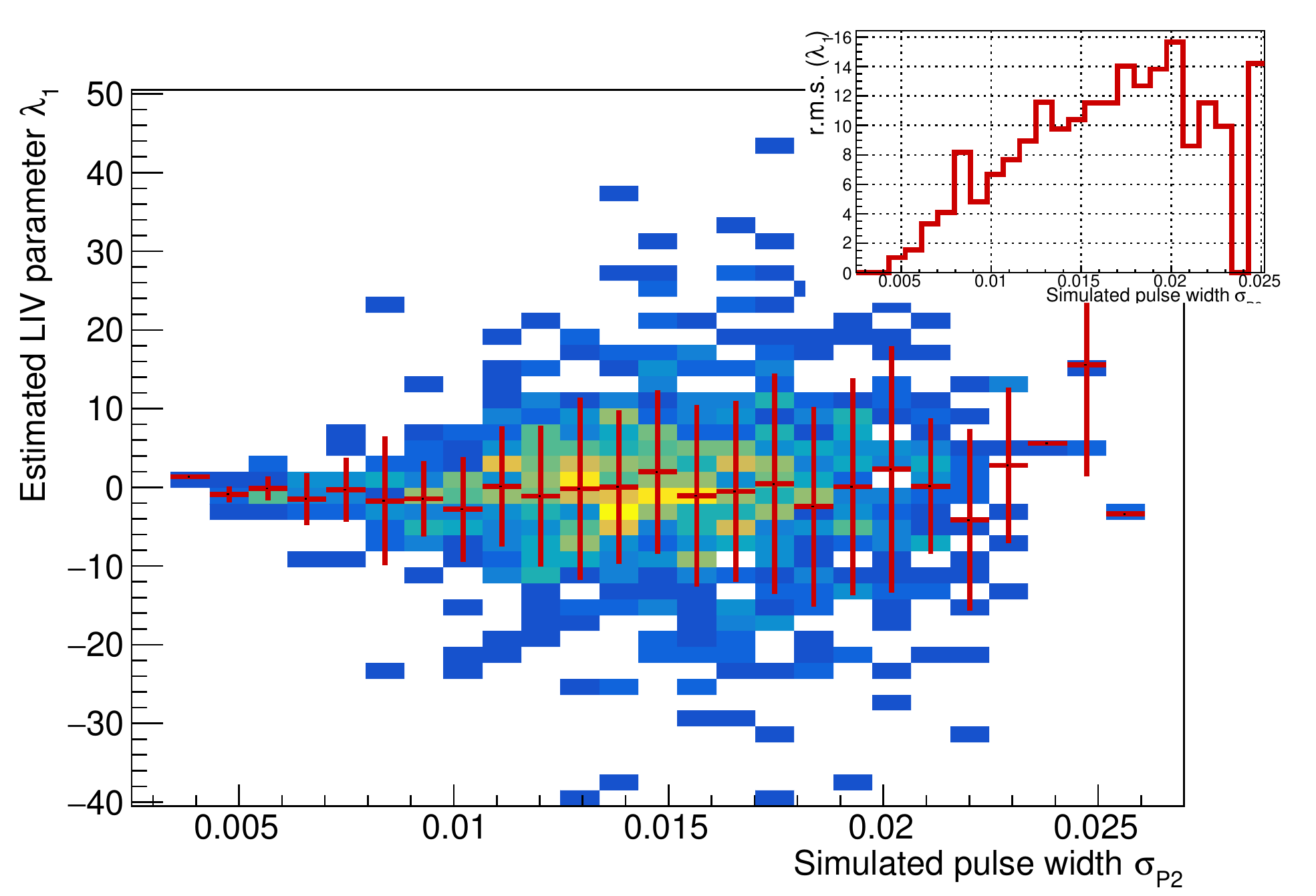} 
\caption{Left: distribution of reconstructed pulse width $\hsigmap$ vs. simulated pulse width $\sigmap$, for a set of 1000 MC simulations involving variation of all
  nuisance parameters, under absence of LIV. Below, the residuals are shown. Right: 
  distribution of reconstructed linear LIV intensities $\widehat{\lambda}_1$ for the same set of simulations, as a function of the simulated pulse width $\sigmap$. 
The inlet shows the rms of $\widehat{\lambda}_1$. \label{fig.lambda1vssigma}}
\end{figure}

The effect of the variations of the nuisance parameter $\sigmap$ 
the spread of $\widehat{\lambda}_1$ is displayed in Figure~\ref{fig.lambda1vssigma}. 
One can see that the nuisance parameter is reconstructed correctly on average, i.e. it shows no bias with respect to the simulated ones. 
The reconstructed LIV parameter $\widehat{\lambda}_1$, although not correlating with the simulated nuisance parameter, shows a reconstruction uncertainty 
that increases with larger values of simulated $\sigmap$.  
This behavior ultimately produces a stronger peaked distribution of $\widehat{\lambda}_1$ with wider tails.

\section{Exponential cutoff limit}
\label{app.cutoff}

We define a test statistic $D$ for $E_b$, based on the profile likelihood method, similar to the one defined in Eq.~\ref{eq.lr1}, and set the LIV parameters $\lambda_{1,2}$ to zero: 
\begin{equation}
D(E_b|{\textit{\textbf X}}) = -2 \ln \left(\frac{\;\mathcal{L}(E_b;\widehat{\widehat\bonu}|{\textit{\textbf X}},\lambda_{1,2}=0)}
                                                  {  \mathcal{L}(\widehat{E}_b; \widehat\bonu    |{\textit{\textbf X}},\lambda_{1,2}=0)}\right) \quad. \label{eq.lreb}
\end{equation}

\begin{figure}[h!]
\centering
 \includegraphics[width=0.495\textwidth]{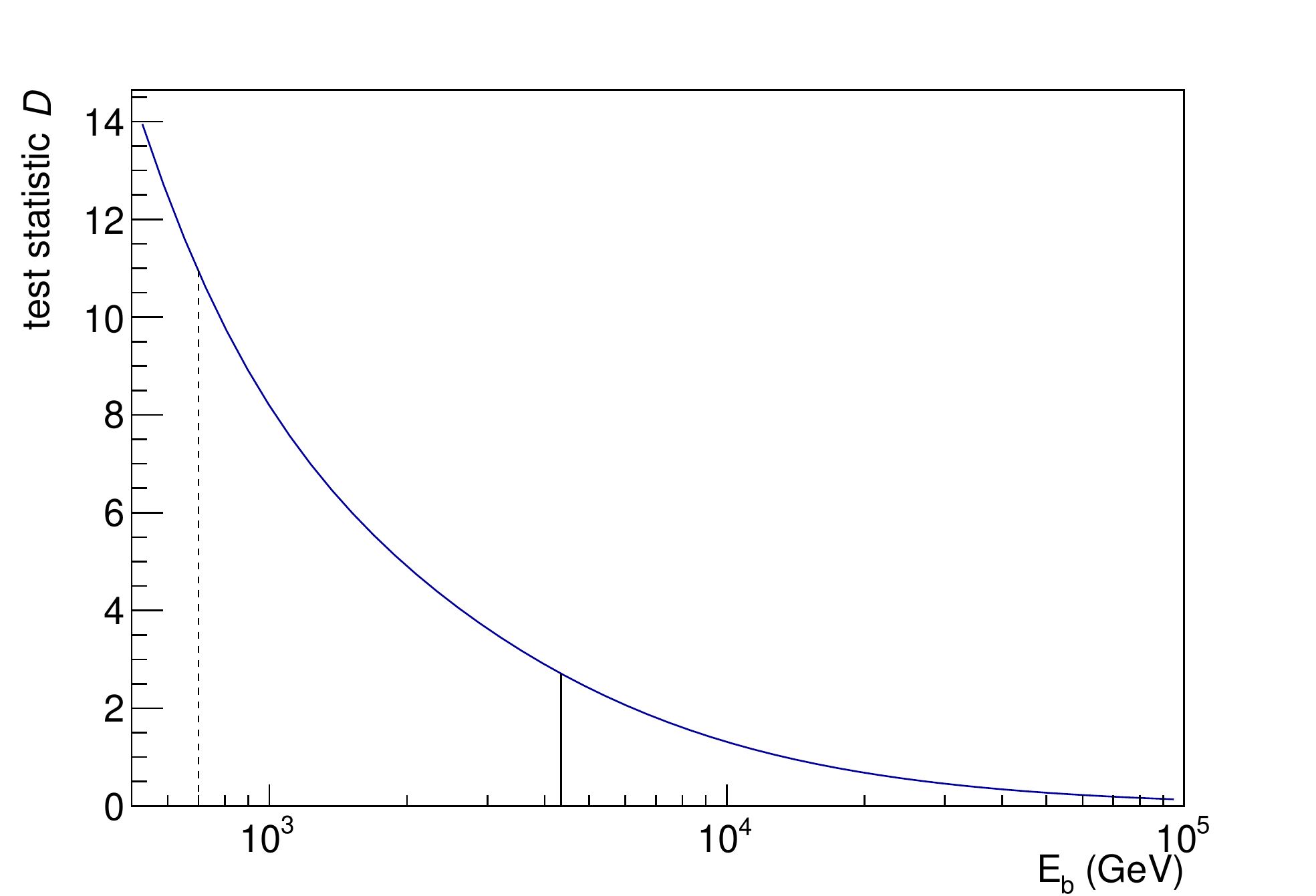}
  \caption{Test statistic (Eq.~\protect\ref{eq.lreb}) as a function of the exponential cutoff energy $E_b$. The full line indicates the energy where $D=2.71$, the dashed line shows 
the previous limit $E_b=700$~GeV.
    \label{fig.lreb}}
\end{figure}

A minimum of $D$ is then found for no cutoff, i.e. $E_b=\infty$. Fig.~\ref{fig.lreb} shows the test statistic as a function of $E_b$. A 95\%~CL limit can be 
evaluated at that value of $E_b$ where $D=2.71$, resulting in $E_b \gtrsim 4.3$~TeV. The value of $D$ is found to be 11.0 at the previous limit $E_b > 700$~GeV (95\%~CL) published in~\citet{crabtev},
who used a binned likelihood approach with fixed nuisance parameters $\phip$ and $\sigmap$.

\end{appendix}



\end{document}